\documentclass[12pt]{article}

\usepackage{amsmath,graphicx,wrapfig,mathtools}
\usepackage{epstopdf} 
\usepackage{multirow}
\usepackage{amsfonts}
\usepackage{amssymb}
\usepackage{amscd}

\usepackage{amsthm,stmaryrd,fancyhdr,latexsym,color,float,fullpage,bm,amscd,tikz-cd,lscape,epsfig,authblk,array,tikz}

\usetikzlibrary{decorations.markings}

\newcommand{\dynkinradius}{.08cm}
\newcommand{\dynkinstep}{.35cm}
\newcommand{\dynkindot}[2]{\fill (\dynkinstep*#1,\dynkinstep*#2) circle (\dynkinradius);}

\newcommand{\dynkinline}[4]{\draw[thin] (\dynkinstep*#1,\dynkinstep*#2) -- (\dynkinstep*#3,\dynkinstep*#4);}

\newenvironment{dynkin}{\begin{tikzpicture}[decoration={markings,mark=at position 0.7 with {\arrow{>}}}]}
{\end{tikzpicture}}

\def\hybrid{\topmargin -20pt    \oddsidemargin 0pt
        \headheight 0pt \headsep 0pt
        \textwidth 6.25in       
        \textheight 9.5in       
        \marginparwidth .875in
        \parskip 5pt plus 1pt   \jot = 1.5ex}

\hybrid

\numberwithin{equation}{section}
\numberwithin{table}{section}

\newcommand{\beq}{\begin{equation}\begin{aligned}}
\newcommand{\eeq}{\end{aligned}\end{equation}}
\newcommand{\bse}{\begin{subequations}}
\newcommand{\ese}{\end{subequations}}
\newcommand{\bi}{\begin{itemize}}
\newcommand{\ei}{\end{itemize}}
\newcommand{\bea}{\begin{eqnarray}}
\newcommand{\eea}{\end{eqnarray}}
\newcommand{\ba}{\begin{array}}
\newcommand{\ea}{\end{array}}
\newcommand{\bt}{\begin{tabular}}
\newcommand{\et}{\end{tabular}}
\newcommand{\bc}{\begin{center}}
\newcommand{\ec}{\end{center}}

\def\nn{\nonumber}

\def\stt{SU(3)_L \times SU(3)_R}

\newcommand{\cref}{{\bf [check ref]}}

\begin{document}


\begin{titlepage}
\begin{center}


\rightline{\small IPhT-T17/057}

\vskip 1.4cm

{\Large \bf  The bosonic string on string-size tori}\\
\vskip 0.4cm

{\Large \bf from double field theory}

\vskip 1.2cm

{\large  Yago Cagnacci${}^{a,b}$, Mariana Gra{\~n}a$^{c}$, Sergio Iguri${}^{a,b}$ and Carmen Nu\~nez${}^{a,b}$ }

\vskip 0.6cm

\small{${}^a${\em  Instituto de Astronom\'ia y F\'isica del Espacio
(CONICET-UBA) and\\
${}^b$Departamento de F\' isica, FCEN, Universidad de Buenos Aires\\
C.C. 67 - Suc. 28, 1428 Buenos Aires, Argentina} \\\
\vskip 0.1cm

{}$^{c}${\em Institut de Physique Th\'eorique,                   
CEA/ Saclay \\
91191 Gif-sur-Yvette Cedex, France}  \\}
\vskip 0.4cm

\end{center}

\vskip 0.6cm

 {\bf Abstract}: We construct the effective action for toroidal compactifications of   bosonic string theory  from generalized Scherk-Schwarz reductions of double field theory. The enhanced gauge symmetry  arising at special points in moduli space is incorporated
 into this framework  by promoting the $O(k,k)$ duality group of $k$-tori compactifications to $O(n,n)$, $n$ being the dimension of the enhanced gauge group, which allows to account for the full massless sector of the theory. We show that the effective action  reproduces the right masses of scalar and vector fields  when moving sligthly away from the points  of maximal symmetry enhancement.   The neighborhood of the enhancement points in moduli space can be neatly explored by spontaneous symmetry breaking. We generically discuss toroidal compactifications of arbitrary dimensions and maximally enhanced gauge groups, and then inspect more closely the example of $T^2$ at the $\stt$ point, which is the simplest setup  containing all the non-trivialities of the generic case.  We show that  the entire moduli space can be described  in a unified way by considering compactifications on  higher dimensional tori.


\noindent

\vfill

\today

\end{titlepage}



\begin{small}
\tableofcontents
\end{small}

\newpage

\section{ Introduction}

The low energy limit of string theory compactifications on string-size manifolds cannot be obtained by usual Kaluza-Klein reductions of ten-dimensional supergravity, since these do not incorporate the light modes originating from strings or branes wrapping cycles of the internal manifold. 
At special points in moduli space, some of these modes  become massless, and there is an enhacement of symmetry promoting the (typically Abelian) gauge groups appearing in Kaluza Klein compactifications to non-Abelian groups. For toroidal compactifications of the heterotic string, this phenomenon has been beautifully described  by Narain \cite{narain}. In compactifications of the bosonic string  on $k$-dimensional tori, the $U(1)_L^k \times U(1)_R^k$ symmetry of the Kaluza-Klein reduction gets enhanced to $G_L \times G_R$ at special points in moduli space, where $G_L, G_R$ are  simply-laced groups\footnote{For simplicity we consider the cases of equal groups on the left and on the right, {\em i.e.} $G_L=G_R=G$, but the construction can be generalized to different groups $G_L, G_R$ of equal rank $k$.} of rank $k$ and dimension $n$, and there are $n^2$ massless scalars  transforming in the $(n,n)$ adjoint representation of the left and right symmetry groups. 
These special points correspond to tori whose radii are of order one in string units, and the states that provide the enhancement have  non-zero winding number besides non-zero momentum in some torus directions.

Capturing winding states within a field theory requires new ingredients beyond those of ordinary Kaluza-Klein compactifications. In the double field theory (DFT) framework \cite{DFT}, these include the introduction of a T-dual coordinate to every torus direction, which is the Fourier dual of the corresponding winding mode (for reviews see \cite{DFTreviews}). 
DFT is therefore a field theory formulated on a double torus incorporating the $O(k,k)$ T-duality symmetry of the bosonic string on a $k$-torus. 
Consistency of the theory requires constraints.  Although the most general form of these constraints is unclear, a sufficient but not necesary constraint is the so-called section condition or strong constraint, that restricts the fields to depend on a maximally isotropic subspace with respect to the $O(k,k)$ inner product, such as the original $k$-torus or its T-dual one, and the original field theory
is recovered but now formulated in an $O(k,k)$-covariant way.  Understanding to what extent this constraint can be violated while keeping a consistent theory remains an open question, about which there are a few answers in particular setups.

The setup relevant to this paper is that of generalized Scherk-Schwarz reductions of double field theory \cite{Arg,geiss} on generalized parallelizable manifolds \cite{waldramspheres}, namely manifolds for which there is a globally defined generalized frame on the double space, such that the C-bracket algebra (the generalization of the Lie algebra that is needed in double field theory) on the frame gives rise to (generalized) structure constants, which are on the one hand trully constant, and on the other should satisfy Jacobi identities. As in standard Scherk-Schwarz reductions \cite{SS}, the only allowed dependence on the internal (double) coordinates is through this frame. Even though the dependence of the frame on internal coordinates might violate the strong constraint, it was shown in \cite{hk,GM} that the theory is consistent at the classical level, as long as the structure constants satisfy the Jacobi identities.   

A generalized Scherk-Schwarz reduction of double field theory on a double circle gives the low energy action for compactifications of the bosonic string  on a circle, and the procedure implemented in   \cite{uscircle} allows to describe the string theory features described above when the radius is close to the self-dual one. In this paper we extend the results of \cite{uscircle}, and show that  the low-energy action for compactifications of the bosonic string on any $k$-torus in a region of moduli space close to a point of symmetry enhancement to a group $G \times G$ can be  obtained from double field theory. To that aim, we consider double field theory on the double torus (of dimension $k+k$), and build an $\frac{O(d+n,d+n)}{O(d+n) \times O(d+n)}$ structure on it (where $d$ denotes the number of external directions and $n$ is the dimension of $G$), given by a generalized metric. The generalized metric, or rather the generalized vielbein for it, is of the Scherk-Schwarz form, namely it is the product of a piece that depends on the external coordinates, and involves the $2n$ vector and $n^2$ scalar fields of the reduced theory that are massless at the enhancement point, and a piece depending on the internal, doubled, $2k$ coordinates. The internal piece is such that the C-bracket algebra gives rise to the $G \times G$ symmetry. Plugging this generalized metric in the double field theory action and following the generalized Scherk-Schwarz reduction of \cite{Arg,geiss}, we obtain an action that exactly reproduces the string theory three-point functions at the point of symmetry enhancement.  Furthermore, we show how the process of symmetry breaking by Higgsing in the effective action gives the exact string theory masses for the vector and scalar fields close to the enhancement point in moduli space, up to second order in  deviations from this point. The Higgsing process amounts to giving vacuum expectation values to the $k^2$ scalars along the Cartan directions of the group $G \times G$, and we show that these vevs are precisely given by departure of the metric and B-field on the torus away from their values at the enhancement point.    

We provide the explicit expression for the generalized vielbein. For $G=SU(2)^k$, the piece of the vielbein that depends on the internal $2k$ coordinates is  a straightforward extension of the one corresponding to $k=1$ that was constructed in \cite{uscircle}. The  $\mathfrak{su}(2)^k_L \times \mathfrak{su}(2)^k_R$ algebra is obtained from the C-bracket of a block-diagonal frame made of $k+k$ $(3 \times 3)$-blocks, where each block involves the vertex operators of the corresponding $SU(2)$ ladder currents. Geometrically, this translates into a 2-dimensional fibration of the directions corresponding to positive and negative roots over the Cartan direction, given by the corresponding circle coordinate $y_{L(R)}=y \pm \tilde y$. The fibration has trivial monodromy. 
For groups that have additionally non-simple roots, 
we show  that the bracket can be deformed in a way that preserves the $O(k,k)$ covariance. The deformation accounts for the cocycle factors that are necessary in the  vertex representation of the current algebra, and then we can reproduce the $\mathfrak{g}\times \mathfrak{g}$  algebra with a generalized vielbein that depends on $k+k$ coordinates only. An alternative generalized frame can be constructed from the formulation of DFT on group manifolds \cite{DFTgroup,dh}, in which it depends on  $n$ coordinates. The question whether there exists a vielbein depending strictly on $k+k$ coordinates that gives rise to the $\mathfrak{g} \times \mathfrak{g}$  algebra  under the usual C-bracket, when $\mathfrak{g}$ has at least one non-simple root, remains open. 

A very interesting question is whether there is a description of the full moduli space, namely a formulation that includes all the states that are massless at any point in the moduli space. We show that such a description requires considering a higher-dimensional torus at a point of maximal enhancement. For  $T^2$ and $T^3$, one gets the effective action at any point in moduli space by considering one of the points of maximal enhancement on a torus of one  higher dimension ($T^3$ and $T^4$ respectively), and combining the process of spontaneous symmetry breaking together with a decompactification limit. For $T^4$, a description that includes the whole moduli space requires considering enhancement points at an even larger torus, namely a $T^7$. We explain how this process works dimension by dimension. 
Note, though, that the action obtained this way does not correspond to a low energy action, since states that are massless at one point in moduli space get string-order masses at another point.

The paper is organized as follows. In section 2 we review toroidal compactifications of  bosonic string theory. We consider the $O(k,k)$ covariant formulation of compactifications on  $T^k$ with constant background metric and antisymmetric 2-form fields and the basics of T-duality. The enhancement of the gauge symmetry at special points in moduli space is discussed in general for $T^k$ and details are provided for the $k=2$ case. The basic features of DFT and generalized Scherk-Schwarz compactifications are reviewed  in section 3. Using this framework, we  construct  the effective action of  bosonic string theory compactified on $T^k$ in the vecinity of a point of symmetry enhancement  in section 4. In particular, we show that a deformation of the C-bracket involving the cocycle factors of the vertex algebra allows to reproduce the structure constants of the enhanced symmetry algebra. In section 5 we check that  the construction reproduces the string theory results when moving slightly away from that point. A higher dimensional formulation that allows  to accommodate all maximal enhancement points in a single approach is presented in section 6. Finally, an overview and conclusions are given in section 7. Three appendices collect the necessary definitions and notation used in the main text. Basic notions of simply laced Lie algebras and Lie groups are reviewed in Appendix A, some basic facts about cocycles are contained in Appendix B and the explicit discussion of symmetry breaking on $T^4$ is the subject of Appendix C.



\section{Toroidal compactification of the bosonic string}
\label{sec:torusstring}

In this section we recall the main features of toroidal compactifications of the bosonic string. We first discuss the generic $k$ case and then we concentrate on the $k=2$  example. For a more complete review see \cite{GPR}.

\subsection{Compactifications on $T^k$}

Consider the bosonic string propagating in a background manifold that is a product of a $d=26-k$ dimensional space-time times an internal torus $T^k$ with a constant background metric
\beq \label{e}
g=e^t e \quad \left( \Rightarrow g_{mn}=e^a{}_m \delta_{ab} e^b{}_n  \right) 
\eeq
 and antisymmetric two-form field $B_{mn}$, $m,n=1,...,k$. For simplicity we take the dilaton to be zero.  The set of vectors $e_m$ define a basis in the compactification lattice $\Lambda^k$ such that the target space is the $k$-dimensional torus $T^k=\mathbb R^k/\pi\Lambda^k$. 

The contribution from the internal sector to the world-sheet action is 
\begin{equation}
\label{action}
S=\frac{1}{4\pi} \int_M d\tau d\sigma \left(\delta^{\alpha \beta} g_{mn} - i\epsilon^{\alpha \beta} B_{mn} \right) \partial_{\alpha}Y^{m}\partial_{\beta}Y^{n}.
\end{equation}
The metric and the $B$-field are  dimensionless\footnote{We will write explicit factors of $\alpha'$ later in the text when they are needed for clarification.}, the world-sheet metric has been gauge fixed to $\delta^{\alpha\beta}$ ($\alpha,\beta=\tau,\sigma$) and the internal string coordinate fields satisfy
\beq \label{periodicityY}
Y^m(\tau,\sigma+2\pi)\simeq Y^m(\tau,\sigma) + 2 \pi w^m\ ,
\eeq  
where $\omega^m\in \mathbb Z$ are the winding numbers and
\begin{equation}
Y^m(z, \bar z)=Y^m_L(z)+Y^m_R(\bar z) \ , \quad z=\exp(\tau + i\sigma) \ , \ \bar z=\exp(\tau - i\sigma) \ ,
\end{equation}
with
\bea
Y^m_L(z)&=& y_{L}^m-\frac{i}{\sqrt 2}p^m_L\,  lnz+\cdots,
\nn\\
Y^m_R(\bar z)&=& y_{R}^m-\frac{i}{\sqrt 2}p^m_R \,  ln\bar z+\cdots,
\eea
 the dots standing for the oscillators contribution. 

The periodicity of the wavefunction requires quantization of the canonical momentum\footnote{The unusual $i$ factor is due to the use of Euclidean world-sheet metric.}
\bea
p_m & = & i \frac{\delta S}{\delta \partial_\tau Y^m}=\left(ig_{mn}\partial_{\tau} Y^{n}+B_{mn}\partial_{\sigma}Y^n\right) \nn \\
& = &\frac{1}{\sqrt 2}g_{mn}(p_L^n+p_R^n)
+\frac{1}{\sqrt 2}B_{mn}(p_L^n-p_R^n)=n_m \in {\mathbb Z}\, ,
\eea
and \eqref{periodicityY} implies the quantisation condition 
\beq
Y^m(\tau,\sigma+2\pi)-Y^m(\tau,\sigma)=(p^m_L-p_R^m)\frac{2\pi}{\sqrt 2}=2\pi \omega^m \ .
\eeq

These equations give 
\bse \label{pa}
\begin{equation}
\label{paL}
p_{aL}=\frac{1}{\sqrt 2}\hat e_a{}^{m}\left[n_m+(g_{mn}-B_{mn}) \omega^n \right], 
\end{equation}
\begin{equation}
\label{paR}
p_{aR}=\frac{1}{\sqrt 2}\hat e_a{}^{m}\left[n_m - (g_{mn}+B_{mn}) \omega^n \right] . 
\end{equation}
\ese
The vectors $\hat e_a$ constitute the canonical basis for the dual lattice $\Lambda^{k*}$, {\em i.e.} $\hat e_a{}^{m} e^a{}_n = \delta^{m}{}_n$, and thus they obey
\begin{equation} \label{eehat2}
\hat e^t \hat e=g^{-1} \quad  \left( \Rightarrow \hat e_a{}^{m} \delta^{ab} \hat e_b{}^{n} = g^{mn} \right) .
\end{equation}
The pairs $(p_{aL},p_{aR})$ transform as vectors under $O(k,k,\mathbb R)$ and they expand the $2k$-dimensional momentum lattice $\Gamma^{(k,k)} \subset {\mathbb R}^{2k}$. From \eqref{pa} one sees they satisfy
\beq
p_L^2-p_R^2=2\omega^m n_m \in 2 \mathbb Z\, ,
\eeq
and therefore they form an even $(k,k)$ Lorentzian lattice. In addition, self-duality  $\Gamma^{(k,k)}=\Gamma^{(k,k)*}$ follows from modular invariance \cite{narain, polchi}. 

The space of inequivalent lattices and inequivalent backgrounds reduces to
\begin{equation}
\frac{O(k,k,\mathbb R)}{O(k,\mathbb R)\times O(k,\mathbb R) \times O(k,k,\mathbb Z) \times \mathbb Z_2}\, ,
\end{equation}
where $O(k,k,\mathbb Z)$ is the T-duality group (we give more details about it in the next section), and the $\mathbb Z_2$ factor accounts for the world-sheet parity $\sigma \rightarrow -\sigma$, a symmetry acting on the background as $B_{mn} \rightarrow -B_{mn}$.

\subsection{$O(k,k)$ covariant formulation}
\label{sec:Okktoro}

The mass of the states and the level matching condition are respectively given by
\begin{subequations}
\begin{equation}
\label{Ham}
m^2=2 (N+\bar N-2) + \left(p_L^2+p_R^2\right),
\end{equation}
\begin{equation}
\label{constraint}
0=2 (N-\bar N) + \left(p_L^2-p_R^2\right).
\end{equation}
\end{subequations}

These can be written in terms of the momentum and winding numbers using an $O(k,k)$-covariant language by introducing the vector $Z$ and  the $O(k,k,\mathbb R)$  invariant metric   $\eta$
\beq \label{Zeta}
Z= \begin{pmatrix} \omega^m \\ n_m \end{pmatrix}  \ , \qquad \eta=\begin{pmatrix} 0 & 1_k \\ 1_k & 0 \end{pmatrix} \ ,
\eeq
as well as the ``generalized metric" of the $k$-dimensional torus, given by the $2k \times 2k$ matrix
\beq
\label{G}
{\cal H}=\begin{pmatrix} g-Bg^{-1}B & Bg^{-1} \\ -g^{-1}B & g^{-1} \end{pmatrix} \  \in  O(k,k,\mathbb R)\, .
\eeq
 
The mass formula \eqref{Ham} and the level matching condition \eqref{constraint} then read
\bse 
\beq \label{MZhZ}
m^2=2 ( N + \bar N -2) +  Z^t {\cal H} Z \ ,
\eeq
\beq \label{LM}
0= 2 (N - \bar N) +  Z^t \eta Z \ ,
\eeq
\ese
respectively.

Note that both the mass formula and the level matching condition are invariant under the T-duality group $O(k,k,\mathbb Z) $
acting as
\beq
Z \to O Z \ , \quad {\cal H}\to O {\cal H} O^t \ , \quad \eta \to O \eta O^t=\eta \ , \quad O\in O(k,k,\mathbb Z) \ .
\eeq
The group $O(k,k,\mathbb Z)$ 
is generated by integer theta-parameter shifts, associated with the addition of an antisymmetric integer matrix $\Theta_{mn}$ to the antisymmetric $B$-field,
\beq \label{theta}
O_{\Theta}=\begin{pmatrix} 1 & \Theta \\ 0 & 1 \end{pmatrix} \ , \quad \Theta_{mn} \in {\mathbb Z}\, ,
\eeq
lattice basis changes 
\beq \label{GLk}
O_{M}=\begin{pmatrix} M & 0 \\ 0 & (M^t)^{-1} \end{pmatrix} \ , \quad M \in GL(k,{\mathbb Z})\, ,
\eeq
and factorized dualities, which are generalizations of the $R\rightarrow 1/R$ circle duality, of the form
\beq \label{Di}
O_{D_i}=\begin{pmatrix} 1-D_i & D_i \\ D_i & 1-D_i \end{pmatrix} \ , 
\eeq
where $D_i$ is a $k \times k$ matrix with all zeros except for a one at the $ii$ component. 

Notice the particular role played by the element $\eta$ viewed as a sequence of factorized dualities in all tori directions, {\em i.e.} 
\beq
\eta=O_D\equiv \prod_{i=1}^k O_{D_i} \ .
\eeq
Its action on the generalized metric is
\beq \label{Tall}
{\cal H} \to O_D {\cal H} O_D^t = \begin{pmatrix} g^{-1} & -g^{-1}B \\  Bg^{-1} & g-Bg^{-1}B \end{pmatrix} = {\cal H}^{-1} \ ,
\eeq
and, together with the transformation $Z \to O_D Z $ which accounts for the exchange $w^m\leftrightarrow n_m$, it generalizes the $R\leftrightarrow 1/R$ duality of the circle compactification. These transformations define the dual coordinate fields (up to the center of mass coordinates)
\bea \label{tildeY}
\tilde Y_m(z,\bar z)&=&-i [(g_{mn}-B_{mp}g^{pq}B_{qn})w^n+B_{mn}g^{np}n_p]\tau + n_m\sigma +\cdots\nn\\
&=& g_{mn}(Y^n_L-Y^n_R) + B_{mn}(Y^n_L+Y^n_R)\, ,
\eea
the dots standing for the oscillator contributions.

A vielbein $E$ for the generalized metric 
\beq \label{E}
{\cal H}=E^t E,
\eeq
can be constructed from the vielbein for the metric \eqref{e} and inverse metric \eqref{eehat2}, as follows
\beq \label{genframe}
E=\begin{pmatrix} E^a \\ E_a \end{pmatrix}= \begin{pmatrix*}[l] \ \ \ \ \ {e}^a  \\   \hat e_a -\iota_{{\hat e}^{{a}}} {B}\end{pmatrix*}=\begin{pmatrix} e & 0 \\ -\hat e B & \hat e
\end{pmatrix}\, .
\eeq
In the basis of left and right movers, that we call ``LR", where the $O(k,k,\mathbb R)$ metric $\eta$ takes the diagonal form
\beq \label{etaC+C-}
\eta_{LR}=(R \eta R^{T})=\begin{pmatrix} 1 & 0 \\ 0 & -1 \end{pmatrix}\, ,  \quad R=\frac{1}{\sqrt2}\left(\begin{matrix}
1 &1
\\
-1& 1\end{matrix}\right)\ ,
\eeq
the vielbein is
\beq \label{genframeLR}
E_{LR}\equiv R E\equiv \begin{pmatrix} E_{aL} \\ E_{aR} \end{pmatrix}=\frac{1}{\sqrt2}\begin{pmatrix} e -\hat e B & \hat e \\ -e -\hat e B & \hat e
\end{pmatrix} \ .
\eeq
Note that this is not the most general parameterisation for the generalized vielbein. We could have used on the first line a vielbein for the (ordinary) metric $e$, and its inverse $\hat e$, and on the second line, which corresponds to the right sector, a different vielbein $\bar e$, $\hat {\bar e}$ giving rise to the same metric $\bar e^t\bar e=e^te=g$. For simplicity we use in most of the text the same vielbein on the left and on the right, except later in section \ref{effectiveaction} (see in particular Eq \eqref{Mab}), where we need to make use of this freedom. 

Then the momenta $(p_{aL}, p_{aR})$ in \eqref{pa} are 
\beq \label{pE}
 \begin{pmatrix} p_{aL} \\ p_{aR} \end{pmatrix} = E_{LR} \, Z \ .
\eeq

\subsection{Gauge symmetry enhancement}
\label{sec:massless}

At special points in moduli space there is an enhancement of gauge symmetry due to the fact that there are extra massless states with non-zero momentum or winding on the torus. From \eqref{Ham} and \eqref{constraint},  the massless states satisfy 
\bse
\begin{equation}
\label{pN}
p_L^2+2N-2=0,
\end{equation}
\begin{equation}
\label{pN2}
p_R^2+2\bar N-2=0,
\end{equation}
\ese
and therefore, $N,\bar N\le 1$. 
This means there are, from the point of view of the non-compact $d$-dimensional space-time, massless 2-tensors (given by the usual states with no momentum or winding and $N_x=\bar N_x=1$)\footnote{$N_x$ ($N_y$) denote the oscillation numbers along the non-compact (compact) directions.}, massless vectors (with $N_x=1, \bar N_x=0$ or $N_x=0, \bar N_x=1$), and massless scalars (with $N_x=\bar N_x=0$).

Let us concentrate on the vectors first, and analyze the case $N_x=0, \bar N_x=1$. There are two types of massless states of this form, those with $N_y=1$ and no momentum or winding, and states with $N_y=0$ and winding or momentum such that $p_L^2=2$, $p_R=0$. The former are the $k$ Kaluza-Klein (KK) vectors generating $U(1)_L^k$ in the left sector, which are massless at any point in moduli space. The fields and their vertex operators are 
 \begin{equation}
\label{vertexcartan}
A_{\mu}^m \to V(z,\bar z) = \varepsilon_{\mu}^m : H^{m}(z)  \bar\partial X^{\mu} \,  e^{iK^{\rho} X_{\rho}}: \ , \quad H^{m}(z) = i \sqrt2\partial Y^{m} = i  \sqrt2\partial Y_L^{m} \, ,
\end{equation}
where $\mu,\nu,\rho=0,1,...,d-1$, and $K^\rho$ is the $d$-dimensional momentum in space-time. The massless states with no oscillation but momentum or winding number along internal directions have vertex operators
 \begin{equation}
\label{vertexroots}
A_{\mu}^\alpha \to V(z,\bar z) = \varepsilon_{\mu}^{\alpha} :  J^{\alpha}(z) \, \bar\partial X^{\mu}  \, e^{iK^{\rho} X_{\rho}}: \ ,\quad J^{\alpha}(z) = c_{\alpha} \tilde J^{\alpha}(z)= c_{\alpha} e^{i \sqrt2\alpha_{m} y_L^m(z)}, \quad 
\end{equation}
where $\alpha_m$ label the compact left-moving momenta, related to \eqref{paL} by 
\beq
\alpha_m  =e^a{}_m \,  p_{aL},
\eeq
and $c_{\alpha}$ is a cocycle introduced to ensure the right properties of the OPE between vertex operators (see Appendix \ref{app:cocycles} for more details).
The reason for introducing new notation for the left-moving momentum will become clear in a moment. The OPE of the currents $J^{\alpha}$ with those in the KK sector \eqref{vertexcartan} is
\beq \label{cartanroot}
H^m(z) J^\alpha (w) \sim \frac{\alpha^m J^\alpha (w)}{z-w} \ ,
\eeq
and allows to identify $H^m$ as the Cartan currents and $J^\alpha$ as the currents corresponding to a root $\alpha$ of the holomorphic part of the enhanced algebra.
The OPE between two $\tilde J^\alpha$ reads
\beq \label{OPEalphabeta}
\tilde J^\alpha(z)\tilde J^\beta(w)&\sim &(z-w)^{(\alpha ,\beta)}\tilde J^{\alpha +\beta}(w)+(z-w)^{(\alpha,\beta)+1}\alpha\cdot i\partial y \tilde J^{\alpha+\beta}(w)+\cdots 
\eeq
where $(\alpha,\beta)=\alpha^m\beta^nK(H^m,H^n)$ denotes  the inner product. The Killing form $K$ is given in \eqref{KillingH}, and is just a $\delta^{mn}$ in the Cartan-Weyl basis. Singular terms appear in the OPE only if $(\alpha,\beta)$ is equal to $-2$ or $-1$. 
If $(\alpha, \beta)=-2$, then $\beta =-\alpha$ and one gets from (\ref{OPEalphabeta})
\bea \label{plusminusroot}
 J^\alpha(z) J^{-\alpha}(w)\sim \frac1{(z-w)^2}+\frac{\alpha\cdot H}{z-w} \ .
\eea
If $(\alpha ,\beta)=-1$, then $\alpha +\beta$ is a root and \eqref{OPEalphabeta} reads 
\bea \label{tworoots}
J^\alpha (z)  J^\beta(w)\sim (-1)^{\alpha \ast \beta} \frac{J^{\alpha+\beta}}{z-w} + ...
\eea
The sign  is determined by the  $\ast$  product defined in Eq. \eqref{ast} and can be reproduced by the cocycles. It
 makes the OPE invariant under $\alpha \leftrightarrow \beta$ and $z \leftrightarrow w$.  

Altogether Eqs \eqref{cartanroot},  \eqref{plusminusroot} and \eqref{tworoots} show that the currents $H^m, J^\alpha$ satisy the OPE algebra of the holomorphic part of the enhanced symmetry  group. 
This group has rank $k$ (the dimension of the torus). Furthermore, the requirement $p_L^2=2$ implies that the enhanced symmetry must correspond necessarily to a simply laced algebra. We denote the dimension of the left-moving part of the algebra by $n$. The number of roots $\alpha$ is equal to $(n-k)$.  
We will see below some examples of enhanced gauge groups.  In Appendix \ref{app:Liealgebras} we collect  the necessary definitions, notation and conventions regarding Lie algebras and Lie groups. 

The right part of the enhanced gauge algebra is constructed in an analogous way from states with $N_x=1, \bar N_x=0$, and either $\bar N_y=1$ and no momentum or winding, or $\bar N_y=0$ and momentum and winding such that $p_L=0, p_R^2=2$. The vertex operators for the former are
\begin{equation}
\label{vertexcartanR}
\bar A_{\mu}^m \to  V(z,\bar z) = \bar\varepsilon_\mu^{m} : \bar H^{m}(\bar z) \, \partial X^{\mu} e^{iK^{\rho} X_{\rho}}: \ , \quad \bar H^{m}(z) = i \sqrt2\bar \partial Y^{m}=i \sqrt2\bar \partial Y_R^{m} \ ,
\end{equation}
and for the latter  
 \begin{equation}
\label{vertexrootsright}
\bar A_{\mu}^\alpha \to  V(z,\bar z) = \bar\varepsilon_\mu^{\alpha}: \bar J^{\alpha}(z) \, \partial X^{\mu} e^{iK^{\rho} X_{\rho}}: \ , \quad \bar{ J}^{\alpha}(\bar z) = c_{\alpha}\bar{\tilde J}^\alpha= c_\alpha e^{i \sqrt2\bar \alpha_{m} Y_R^m(\bar z)} 
\end{equation}
with
\beq
\bar \alpha_m=e^a{}_m \,  p_{aR} 
\eeq
a root of the Lie group corresponding to the right part  of the enhanced gauge symmetry. This group has rank $k$ as well, but might not be the same as the one on the left. For simplicity, we will consider from now on the case of equal groups on the left and on the right, which is what happens at points of maximal enhancement (to be discussed later) that will be our primary focus. Almost all of the formulas have though a straighforward generalisation to a gauge symmetry  enhancement group $G_L \times  G_R$ of rank $(k,k)$ and dimension $(n_L, n_R)$. 

Regarding the extra massless scalars, one gets a total of $n^2$. $k^2$ of them are the usual KK scalars with $N_y=\bar N_y=1$ and no momentum or winding, corresponding to the internal metric and $B$-field. The fields and their vertex operators are
\begin{equation}
\label{vertexcartanscalars}
M_{mn} \to V(z,\bar z) =\varepsilon_{mn} :  H^{m}(z) \bar H^n(\bar z) e^{iK^{\rho} X_{\rho}}: \ .
\end{equation}
They are massless at any point in moduli space, and their vev's determine the type of symmetry enhancement. At the points in moduli space where the symmetry is enhanced to a group of dimension $(n,n)$, there are $(n-k) \times k$ scalars with $N_y=0, \bar N_y=1$, $p_L^2=2$ and vertex operators
\begin{equation}
\label{vertexGoldstoneleft}
M_{\alpha n} \to V(z,\bar z) =\varepsilon_{\alpha n} : J^{\alpha}(z) \bar H^{n}(\bar z) \, e^{iK^{\rho} X_{\rho}}: \ ,  
\end{equation}
as well as $k \times (n - k)$ scalars with $\bar N_y=0,  N_y=1$, $p_R^2=2$ and vertex operators
\begin{equation}
\label{vertexGoldstoneright}
M_{m \beta} \to V(z,\bar z) = \varepsilon_{m \beta}:  H^{m}(z) \bar J^{\beta}(\bar z) \, e^{iK^{\rho} X_{\rho}}: \ .
\end{equation}
As we will see, in the effective theory around a point in moduli space where there is symmetry enhancement, these are the Goldstone bosons in the symmetry breaking process.  Finally, there are $(n-k)^2$ scalars with no oscillation number and $p_L^2=p_R^2=2$. Their vertex operators are
\begin{equation}
\label{vertexscalars}
M_{\alpha\beta} \to  V(z,\bar z) =\varepsilon_{\alpha\beta} : J^{\alpha}(z) \bar J^{\beta}(\bar z) \, e^{iK^{\rho} X_{\rho}}: \ .
\end{equation}

Let us now see explicitly how to find the extra massless vectors and scalars with momentum or winding number.  Their existence depends on the location in moduli space, $i.e.$  it depends on $g_{mn}$ and $B_{mn}$. The massless vectors in the left moving sector should have $p_R=0$, $p_L^2=2$, and therefore satisfy 
\beq
n_m=(g_{mn}+B_{mn}) \omega^n \ , \quad \omega^mg_{mn}\omega^n=1 \ , \quad n_m\omega^m=1 \ .
\eeq
The simplest case to analyse is that of a torus with diagonal metric and all the radii at the self-dual point, together with vanishing $B$-field $(g_{mn}=\delta_{mn}$, $B_{mn}=0$.) For each torus direction there are two extra massless vectors with $n_m=\omega^m=\pm 1$. These combine with the KK vectors (\ref{vertexcartan}) to enhance the symmetry from $U(1)_L^k$ to $SU(2)_L^k$. Combining with the right moving sector, one has $SU(2)^k_L \times SU(2)^k_R$. 
Other examples of symmetry enhancement groups are found at points in moduli space which are fixed points of a subgroup of the $O(k,k,\mathbb Z)$  T-duality group. 

Maximal enhancement\footnote{``Maximal'' stands here for an enhanced semi-simple and simply-laced symmetry group of
rank $k$ corresponding to a level 1 affine Lie algebra} occurs when the background is a fixed point of the symmetry \eqref{Tall} up to an identification by a theta shift (\ref{theta}) and an $SL(k,{\mathbb Z})$ transformation \eqref{GLk}, namely when 
\beq \label{enhcond}
{\cal H}^{-1} = O_{M} O_{\Theta} {\cal H} O_{\Theta}^t O_{M}^t\, .
\eeq  
The case ${\cal H}=1_{2k}$ just discussed is the simplest one (with $O_M=O_{\Theta}=1$), but there are more general examples in which the background is 
\cite{GRV}
\begin{equation}
\label{metricenhancement}
g_{mn}=\frac{1}{2}A^{mn} \ , \quad 
B_{mn}=\frac{1}{2}\,\mbox{sgn}(m-n)A^{mn},
\end{equation}
where $\mbox{sgn}$ denotes the sign function and $A^{mn}$ is the Cartan matrix associated to the corresponding algebra. Note that the matrices $g+B$ and its transpose $g-B$ at the enhancement point acquire a triangular form. 
Non-maximal enhanced symmetries can be found at fixed points of factorized dualities instead of the full inversion transformation \eqref{Tall}.

We will analyse in the next section the case of $T^2$ compactifications  in detail, where the gauge groups of maximal enhancement are $SU(2)^2_L \times SU(2)^2_R$ and $SU(3)_L \times SU(3)_R$.

\subsection{Compactifications on $T^2$}

Here we discuss in detail the $k=2$ case. The $k^2=4$ moduli can be joined conveniently into two complex fields as follows. The complex structure is given by
\begin{equation}
\tau=\tau_1+i\tau_2=\frac{g_{12}}{g_{22}}+i\frac{\sqrt{g}}{g_{22}},
\end{equation}
while the K\"ahler structure is introduced as
\begin{equation}
\rho=\rho_1+i\rho_2=B_{12}+i\sqrt{g},
\end{equation}
where $g=g_{11}g_{22}-g_{12}^2$ is the determinant of the metric on the torus. The inverse relations read
\begin{equation} \label{gBtaurho}
g=
\frac{\rho_2}{\tau_2}\left(\begin{array}{cc}
	|\tau|^2 & \tau_1 \\
	\tau_1 & 1
\end{array}\right) \ , \qquad 
B=
\left(\begin{array}{cc}
	0 & \rho_1 \\
	-\rho_1 & 0
\end{array}\right).
\end{equation}
Later we will need a vielbein for the metric and its inverse, which can be taken to be
\begin{equation}
e=\sqrt{\frac{\rho_2}{\tau_2}}\left(
\begin{array}{cc}
	\tau_2 & 0 \\
	\tau_1 & 1
\end{array}
\right) \ , \qquad \hat e\equiv (e^{-1})^t =\frac{1}{\sqrt{\rho_2\tau_2}}\left(
\begin{array}{cc}
	1 & -\tau_1 \\
	0 & \tau_2
\end{array}
\right).
\end{equation}

The generalized metric, defined in \eqref{G}, reads in terms of $\tau$ and $\rho$
\beq \label{GT2}
{\cal H}=\frac{1}{\rho_2 \tau_2}\left(
\begin{array}{cccc}
|\rho|^2|\tau|^2 &|\rho|^2 \text{$\tau_1$} & -\text{$\rho_1$} \text{$\tau_1$} & \text{$\rho_1$}|\tau|^2 \\
|\rho|^2 \text{$\tau_1$} &  |\rho|^2 & -\text{$\rho_1$} & \text{$\rho_1$} \text{$\tau_1$} \\
 -\text{$\rho_1$} \text{$\tau_1$} & -\text{$\rho_1$} & 1 & -\text{$\tau_1$} \\
 \text{$\rho_1$}|\tau|^2 & \text{$\rho_1$} \text{$\tau_1$} & -\text{$\tau_1$}& |\tau|^2 \\
\end{array}
\right)\, ,
\eeq
and the corresponding generalized veilbein \eqref{genframe} is
\beq \label{ET2}
E=\frac{1}{\sqrt{\rho_2 \tau_2}}\left(
\begin{array}{cccc}
 \text{$\rho_2$} \text{$\tau_2$} & 0 & 0 & 0 \\
 \text{$\rho_2$} \text{$\tau_1$} & \text{$\rho_2$} & 0 & 0 \\
 -\text{$\rho_1$} \text{$\tau_1$} & -\text{$\rho_1$} & 1 & -\text{$\tau_1$} \\
 \text{$\rho_1$} \text{$\tau_2$} & 0 & 0 & \text{$\tau_2$} \\
\end{array}
\right)\, .
\eeq

The left and right moving momenta \eqref{pa}  in terms of $\tau$ and $\rho$ read\footnote{Here we are expressing the vectors $p_{aL}$ and $p_{aR}$ on the tangent space of $T^2$ as complex variables.}    
\begin{subequations}
\beq
p_L=\frac{1}{\sqrt{2\rho_2 \tau_2}} \left[(n_1-\bar \tau n_2) - \bar \rho (\omega^2+\bar \tau \omega^1) \right]\ , 
\eeq
\beq
p_R=\frac{1}{\sqrt{2\rho_2 \tau_2}} \left[ (n_1-\bar \tau n_2) -  \rho (\omega^2+\bar \tau \omega^1) \right] \ .
\eeq
\end{subequations}

The moduli space is isomorphic to $O(2,2,\mathbb R)/(O(2,\mathbb R) \times O(2,\mathbb R)) =SL(2,\mathbb R)/U(1)\times SL(2,\mathbb R)/U(1)$, where $\tau$ and $\rho$ sweep each $SL(2,\mathbb R)/U(1)$ factor, respectively.
The duality group is generated by the  usual $S$ and $T$ modular transformations, together with the factorized duality $D$ exchanging the complex and the K\"ahler structures\footnote{\label{foot:ST}In terms of the $O(k,k,{\mathbb Z})$ transformations given in the previous section, $S$ and $T$ are the $GL(2)$ transformations in (\ref{GLk}) given respectively by $M_S=\begin{pmatrix} 0 & -1 \\ 1 & 0 \end{pmatrix}$ and $M_T=\begin{pmatrix} 1 & 1 \\ 0 & 1 \end{pmatrix}$, while $D$ is a T-duality transformation $O_{D_2}$. }
\begin{equation} \label{ST}
S: (\tau,\rho) \longrightarrow \left(-1/\tau,\rho\right) \ , \quad
T: (\tau,\rho) \longrightarrow \left(\tau+1,\rho\right)\ , \quad D: (\tau,\rho) \longrightarrow \left(\rho,\tau\right) \ .
\end{equation}
Worldsheet parity acts by
\begin{equation} \label{W}
W: (\tau,\rho) \longrightarrow \left(\tau,-\bar \rho\right).
\end{equation}   
The fundamental domain is given by two copies of the domain shown in Figure 1. 
\begin{figure} \label{fig:funddomain}
\begin{center}
\includegraphics[scale=0.7]{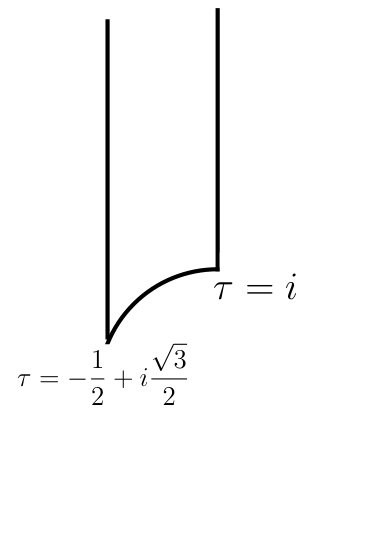}
\end{center}
\vskip -1.7cm
\caption{Fundamental domain for the modulus $\tau$.}
\end{figure}

The possible groups of symmetry enhancement for the left or right sector have rank 2, and are thus $U(1) \times U(1)$ (no enhancement), $SU(2) \times U(1)$ or the maximal $SU(2) \times SU(2)$ or $SU(3)$. All of these occur on the plane $\tau=\rho$  (up to identifications under the discrete symmetries \eqref{ST} and \eqref{W}). As discussed in the previous subsection, enhancement to $SU(2)^2_L \times SU(2)^2_R$ occurs in a compactification with metric given by the identity ($i.e.$ all radii equal to the string length) and no $B$-field. This satisfies \eqref{metricenhancement}, where the Cartan matrix for $SU(2)^2$ is given in \eqref{ASU22}, and corresponds to $\tau=\rho=i$. The Cartan matrix of $SU(3)$ is given in \eqref{ASU3}, and thus according to
\eqref{metricenhancement}, the $SU(3)_L \times SU(3)_R$ enhancement point is reached for $B_{12}=-\tfrac12$, $g_{11}=g_{22}=1$, $g_{12}=-1/2$. This corresponds to  $\tau=\rho=-\tfrac12 + i \tfrac{\sqrt{3}}{2}$, which is at the other corner of the fundamental domain in Figure 1. We will discuss the physics around these two points in detail in the next subsections. $(SU(2) \times U(1))_L \times (SU(2) \times U(1)_R$ enhancement occurs at the borders of the fundamental region, namely 
at $\rho=\tau=-1/2+i \tau_2$, $\rho=\tau=i \tau_2$ and $|\rho|=|\tau|=1$.  At the interior of the region, the enhancement group is $U(1)^2_L \times (SU(2) \times U(1))_R$. This asymmetry between the left and right sectors can be understood from the fact that points at the interior are not fixed points of the ${\mathbb Z}_2$ symmetry $B\to -B$ (or $\rho_1\to -\rho_1$). The mirror region, which  is to the right of the region displayed in Figure 1  in our conventions, has enhanced gauge symmetry group $(SU(2) \times U(1))_L \times  U(1)^2_R$. 

The locations of these groups on the domain $\rho = \tau$ are displayed in Figure 2. 

\begin{figure} \label{fig:enhancements}
\begin{center}
\includegraphics[scale=0.7]{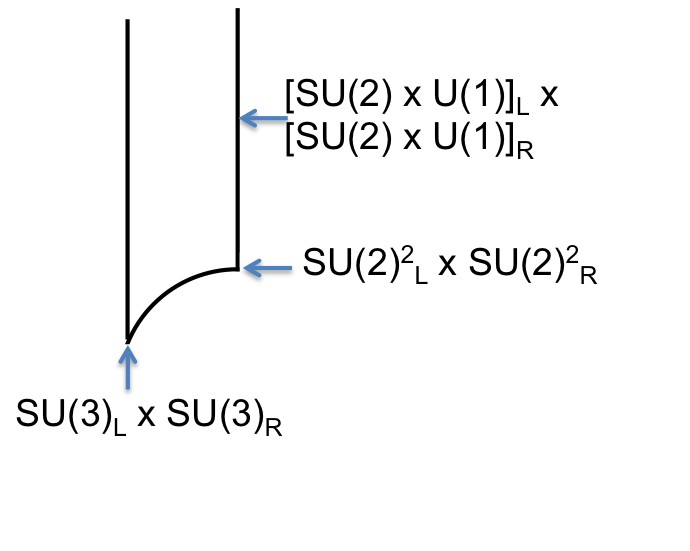}
\end{center}
\vskip -1.7cm
\caption{Enhancement groups on the domain $\tau=\rho$. The groups $SU(3)_L\times SU(3)_R$ and $SU(2)^2_L\times SU(2)^2_R$ occur at the isolated points on the left and right vertices, respectively, while $[SU(2)\times U(1)]_L\times [SU(2)\times U(1)]_R$ occurs along the boundaries. }
\end{figure}

\subsubsection{$SU(2)^2_L \times SU(2)^2_R $ enhancement point} 
 
At $\tau=\rho=i$ there is a gauge symmetry enhancement point with  $SU(2)^2_L \times SU(2) ^2_R$ gauge group. This corresponds to the two-torus being a product of two circles at the self-dual radii, namely the metric is given by the identity and there is no $B$-field.  This satisfies $g_{mn}=\frac{1}{2}A_{mn}$, where $A_{mn}$ is the Cartan Matrix of the  $\mathfrak{su}(2)^2$ algebra given in \eqref{ASU22}.
We choose the vectors $\hat e^{i}$ defined in (\ref{eehat2}) to be 
\begin{align}
\label{vectoresduales}
\hat e_{1}&=(1,0), \qquad
\hat e_{2}=(0,1)\, .
\end{align}


At this point there are $12$ massless vectors and $36$ massless scalars. The left vectors associated to the Cartan subalgebra are given by (\ref{vertexcartan}) and those associated to the ladder operators by (\ref{vertexroots}), where now the $\alpha$ in (\ref{vertexroots}) are the roots of the  $\mathfrak{su}(2)^2 $ algebra.
For the left gauge group, the massless ladder vectors  at $\tau = \rho= i$ are those in Table 2.1. There is an identical construction for the right sector, with appropiate $n_i$ and $w^i$.

\begin{table} 
\begin{center}
\begin{tabular}{|cccc|c|c|c|c|}
\hline	
$n_1$ & $n_2$ & $w^1$ & $w^2$ & $\sqrt{2}\;p_{m,L} $ &$\sqrt{2}\;p_{m,R} $ & root  & vector \\ \hline	
 1 & 0 & 1 & 0 & (2,0) & (0,0) &  $\alpha_{1}$ & $A_L^{\underline 1} $\\
 0 & 1 & 0 & 1 & (0,2) & (0,0) &   $\alpha_{2}$ & $A_L^{\underline 2} $ \\
 1 & 0 & -1 & 0 & (0,0) & (2,0) &   $\alpha_{1}$ & $A_R^{\underline 1}$\\
 0 & 1 & 0 & -1 & (0,0) & (0,2) &   $\alpha_{2}$ & $A_R^{\underline 2}$ \\
 \hline
\end{tabular}
\caption{Massless vectors with momentum and winding at the $SU(2)^2_L \times SU(2)^2_R$ enhancement point $(\tau, \rho)=(i,i)$. Only those associated with positive roots are shown. }\label{ta:masslessSU2}
\end{center}
\end{table}

\subsubsection{$SU(3)_L \times SU(3)_R$ enhancement point}
\label{sec:stt}
%

At $\tau=\rho=-\tfrac{1}{2}+i \tfrac{\sqrt{3}}{2}$ there is a symmetry enhancement point. The resulting gauge group is $SU(3)_L \times SU(3)_R$. The metric and $B$-field are given by
\begin{equation}
\label{gsu3}
g_{mn}=\frac12 \left(
\begin{array}{cc}
 2 & -1 \\
 -1 & 2 
\end{array}
\right),
\qquad
B_{mn}=\frac12 \left(
\begin{array}{cc}
 0 & -1 \\
 1 & 0 
\end{array}
\right),
\end{equation}
where $g_{mn}=\frac{1}{2}A_{mn}$,  $A_{mn}$ being the Cartan matrix of the $\mathfrak{su}(3)$ algebra.
We choose the vectors $\hat e_{a}$ defined in (\ref{eehat2}) to be
\begin{align}
\label{vectoresduales}
\hat e_1=(2/\sqrt3 , 1/\sqrt3) \, ,\qquad  \hat e_2=(0,1)\, .
\end{align}
The generalized metric \eqref{G} (which  is given in \eqref{GT2} in terms of $\tau$ and $\rho$) is 
\beq
{\cal H}= \frac13 \begin{pmatrix} 4 & -2 & -1 & -2   \\ -2 & 4 & 2 &  1 \\  -1 & 2 & 4 & 2\\ -2 & 1 & 2 & 4 
\end{pmatrix} \ .
\eeq
This satisfies \eqref{enhcond} with $O_\Theta=1$ and $O_M=O_{M_S}$ where $M_S$ is defined in footnote \ref{foot:ST}.

There are $16$ massless vectors and $64$ massless scalars at this point in moduli space. The left vectors associated to the Cartan subalgebra are given by (\ref{vertexcartan}) and those associated to the ladder operators by (\ref{vertexroots}), where now  $\alpha=p_{aL}$  are the roots of the  $\mathfrak{su}(3)$ algebra. 
For the left gauge group, the ladder vectors that are massless at the SU(3) point are those in Table 2.2. There is a similar construction for the right sector, with appropiate $n_i$ and $w^i$. In Table 2.3 we give some of the 64 massless scalars, to be used in section \ref{sec:T2su3}.  

Note that the vector $A^{\underline 2}$ in Table \ref{ta:masslessSU3} is not the same as the one denoted $A^{\underline 2}$ in Table \ref{ta:masslessSU2} corresponding to the $SU(2)^2$ case, as the notation refers to the roots of each algebra. Comparing the two tables, we can see that there is an overlap between them and this will be important in section \ref{sec:Td}.

\begin{table}
\begin{center}
\begin{tabular}{|cccc|c|c|c|c|}
\hline	
$n_1$ & $n_2$ & $w^1$ & $w^2$ & $\sqrt{2}\;p_{m,L}$ &$\sqrt{2}\;p_{m,R}$ & root & vector  \\ \hline	
 1 & 0 & 1 & 0 & (2,-1) & (0,0) &   $\alpha_1$& $A_L^{\underline 1} $ \\
 -1 & 1 & 0 & 1 & (-1,2) & (0,0) &   $\alpha_2$& $A_L^{\underline 2} $ \\
 0 & 1 & 1 & 1 & (1,1) & (0,0) &   $\alpha_3$& $A_L^{\underline 3} $ \\
  1& -1 & -1 & 0 &  (0,0) & (2,-1)&   $\alpha_1$& $A_R^{\underline 1}$\\
0  & 1 & 0 & -1 &  (0,0) & (-1,2)&   $\alpha_2$& $A_R^{\underline 2}$ \\
1 & 0 & -1 & -1 &  (0,0) & (1,1)&  $\alpha_3$& $A_R^{\underline 3}$\\
\hline
\end{tabular}
\caption{Massless vectors with momentum and winding at the $SU(3)_L\times SU(3)_R$ enhancement point $\rho=\tau=-1/2+i\sqrt{3}/2$. Only those associated with positive roots are shown.} \label{ta:masslessSU3}
\end{center}
\end{table}

\begin{table} 
\begin{center}
\begin{tabular}{|cccc|c|c|c|c|}
\hline	
$n_1$ & $n_2$ & $w^1$ & $w^2$ & $\sqrt{2}\;p_{m,L}$ &$\sqrt{2}\;p_{m,R}$ & scalar  \\ \hline	
 2 & -1 & 0 & 0 & (2,-1) & (2,-1) & $M_{\underline 1 \underline 1}$   \\
 1& -2 & 0 & 0 &  (1,-2) & (1,-2)& $M_{\overline 2 \overline 2}$ \\
 1& 1 & 0 & 0 &  (1,1) & (1,1)& $M_{\underline 3 \underline 3}$ \\
 1 & 1 & 1 & -1 & (2,-1) & (-1,2) & $M_{\underline 1 \underline 2}$  \\
 2 & -2 & -1 & -1 & (1,-2) & (2,-1) &  $M_{\overline 2 \underline 1}$  \\
 2& -1 & -1 & -2 &  (1,-2) & (1,1)& $M_{\overline 2 \underline 3}$ \\
 \hline
\end{tabular}
\caption{Some of the massless scalars with momentum and winding at the $SU(3)_L\times SU(3)_R$ enhancement point.} \label{ta:masslessscalarsSU3}
\end{center}
\end{table}


\section{Scherk-Schwarz reduction in double field theory}
\label{sec:DFT&SS}

In this section we review Scherk-Schwarz reductions \cite{SS} of double field theory \cite{Arg,geiss}. We start with some very basic notions of double field theory \cite{DFT, OHZ} (for details see for example the reviews \cite{DFTreviews}).  

\subsection{Basics of double field theory}
\label{sec:DFT}

Double field theory incorporates the T-duality symmetry of string toroidal compactifications  in a theory of fields propagating on a double space.  The theory is covariant under a global $O(D,D,{\mathbb R})$ symmetry\footnote{Here $D=d+k$ is the dimensionality of space-time, with $d$ external directions and $k$ internal ones. This however need not always be the case, the $O(D,D)$ covariance might not be associated to $D$ physical dimensions, as we will see in section \ref{sec:enhancementDFT}.}. The propagating degrees of freedom are  the generalized metric \eqref{G} and the dilaton $e^{-2d}=\sqrt{g_D}\, e^{-2\phi}$.
The action is \cite{OHZ}
\beq \label{SDFT}
S=\frac{1}{2k_D^2}\int dX\ e^{-2d} \left( \frac18  {\cal H}^{  M  N}\partial_{  M} {\cal H}^{ K
L}\partial_{  N}  {\cal H}_{  K L}-\frac12 {\cal H}^{ M N}\partial_{
M} {\cal H}^{ K L}
\partial_{  K}  {\cal H}_{  N  L} + ... \right) \ ,
\eeq
where $...$ involve terms with derivatives of the dilaton. The index $M=1,...,2D$ runs over the fundamental representation of $O(D,D)$,  and the doubled coordinates are $X^M=(x^{\mu}, \tilde x_\mu, y^m,\tilde y_m)$, where  $y^m$ are associated to the CFT fields $Y^m$ and their dual $\tilde y_m$ are associated to the dual fields $\tilde Y_m$ in \eqref{tildeY}.

Despite the non-covariant appearence of the action, one can show that it is a scalar under generalized diffeomorphisms\footnote{For this one should  take into account  the terms involving derivatives of the dilaton  in \eqref{SDFT} and recall that $e^{-2d}$ transforms as a density.} \cite{WaldramGG,DFTbracket} 
\bea \label{GL}
({\cal L}_V W)^{M} =V^P\partial_P W^{M}  
- W^{P} \partial_PV^M + \partial^MV_P W^{P} \  \ .
\eea
The algebra of generalized diffeomorphisms closes under the so-called weak and strong contraints
\beq \label{wcsc}
\partial_M\partial^M\cdots =0 \   , \quad \partial_M\cdots \partial^M\cdots =0\, ,
\eeq
where the dots represent any field or gauge parameter. These constraints imply that the fields only depend on half  of the coordinates. The weak constraint is the operator version of the level matching condition \eqref{LM} restricted to the sector $N=\bar N$, namely $Z \to \partial_M$, $Z^t \eta Z \to \partial_M \partial^M$.  Here we also consider states with $N\neq \bar N$ and therefore we will not enforce the weak constraint on the internal double torus. Moreover we will solve the constraints in space-time so that the fields effectively depend on $X^M=(x^\mu, y^m, \tilde y_m)$. The algebra will nevertheless close, as we explain shortly. 

Given a frame $E_A$ for the generalized metric \eqref{E}, one defines the generalized fluxes through
\beq \label{FABC}
{\cal L}_{E_A} E_B=f_{AB}{}^{C} E_C \ .
\eeq 
It is not hard to show that ${f}_{ABC}={f}_{AB}{}^{D} \eta_{DC}$ are totally antisymmetric. 

In this paper we restrict the dependence of the fields on the internal coordinates as done in a Scherk-Schwarz reduction (see \eqref{gss} below), and demand the generalized fluxes with internal indices to be constant. For this particular case, closure of the algebra of diffeomorphisms requires weaker constraints than (\ref{wcsc}), namely the algebra closes as long as the generalized fluxes satisfy the quadratic constraints of gauged supergravity \cite{GM}
\beq \label{quadrconstr}
{f}_{[AB}{}^D {f}_{C]D}{}^E=0\, .
\eeq

\subsection{Generalized Scherk-Schwarz reductions}
\label{sec:SS}

We consider the generalized Scherk-Schwarz reduction of the DFT action \eqref{SDFT} as in \cite{Arg,geiss}. In these reductions, the $O(D,D,{\mathbb R})$ covariance is broken into $GL(d) \times O(k,k,{\mathbb R})$, namely the double tangent space splits into a double tangent space of an internal ``twisted double torus" of dimension $2k$, and a double tangent space for the $d$-dimensional external space, where the $O(d,d)$ covariance is finally broken to $GL(d)$. In this section we use $\hat A=1,...,2D$ as the fundamental (flat) index of $O(D,D)$, and for any vector $V$ this  splits into $V_{\hat A}=(V^\alpha,V_A,V_\alpha)$, where $\alpha=0,...,d-1$ is a flat $d$-dimensional external-space-time index \footnote{Note that the index $\alpha$ is used in all other sections to denote a root of a Lie algebra. There should be no confusion as we need to write explicitly a flat space-time index only in this section, and here we do not need an index for the roots.} and $A=1,...,2k$ runs over the internal twisted double torus. 

For generalized Scherk-Schwarz compactifications, the generalized vielbein is a product of two pieces, one depending on the $d$ external coordinates $x$ and the other one depending on the internal ones, $y$ and $\tilde y$:
\beq \label{gss}
\mathcal E_{\hat A}(x, y, \tilde y)= {\cal U}_{\hat A}{}^{\hat A'}(x) E_{\hat A'}(y, \tilde y) \ .
\eeq
After integrating over the internal double torus, all the information about the internal space will be encoded  in the (constant) generalized fluxes \eqref{FABC}. The matrix ${\cal U}$ parameterises the scalar, vector and tensor fields of the reduced $d$-dimensional action, namely a vielbein for the $d$-dimensional metric $\tilde e$ (by which we mean $\tilde e^{\alpha}{}_\mu (x)$, and its inverse $\hat{\tilde e}_{\alpha}{}^\mu$ will be denoted $\hat{\tilde e}$), a two-tensor $\tilde B=\frac12 \tilde B_{\mu \nu} (x) dx^{\mu} \wedge dx^{\nu}$, $2k$ vector fields $A_{\mu}^A(x) dx^\mu$ and $n^2$ scalar fields encoded in $\Phi_A{}^B(x)$. One can take the following ansatz for the vielbein \cite{geiss} 
\beq \label{gssexplicit}
\mathcal E=\begin{pmatrix} {\mathcal E}^{\alpha} \\ {\mathcal E}_{A} \\ {\mathcal E}_{\alpha}\end{pmatrix}= \begin{pmatrix} \tilde e & 0 & 0 \\
\Phi\cdot A & \Phi & 0 \\
-\hat {\tilde e} (\tilde B+\frac12 A \cdot A) & - \hat {\tilde e} A & \hat{\tilde e} \end{pmatrix} 
\begin{pmatrix} 1 & 0 & 0 \\ 0 & E & 0 \\ 0 & 0 & 1 \end{pmatrix}\, ,
\eeq
where $\cdot$ denotes the $O(k,k)$ internal product, namely $(\Phi \cdot A)_A=\Phi_A{}^B \eta_{BC} A^C$ with $\eta$ the invariant metric, and $E$ here refers to $E_A{}^M$, the $2k \times 2k$ piece of the veilbein depending on the internal coordinates that parameterises the coset $O(k,k,\mathbb R)/O(k)\times O(k)$. 
Choosing $E_A{}^{M}$ as the internal vielbein characterizing the background for the torus, $\Phi_{A}{}^{B}$ describes the fluctuations over this background.

Let us concentrate on the internal part of the vielbein
\beq
 \mathcal E_{A}{}^{M}(x,y,\tilde y)=\Phi_A{}^B(x) E_B{}^M(y, \tilde y) \ . 
 \eeq
 This piece of the frame forms the ``internal generalized metric", defined as\footnote{In case $E_A{}^M$ is complex, this expression should read $\mathcal H^{MN}= \mathcal H^{AB} E_{A}{}^{M*} E_{B}{}^{N}$.}
\begin{equation} \label{HMHA}
\mathcal H^{MN}=  \delta^{AB} \mathcal E_{A}{}^{M} \mathcal E_{B}{}^{N}= \mathcal H^{AB} E_{A}{}^{M} E_{B}{}^{N},
\end{equation}
where
\begin{equation}
\label{HAB}
\mathcal H^{AB}=\delta^{CD} \Phi_{C}{}^{A} \Phi_{D}{}^{B} \ .
\end{equation}

Using the Scherk-Schwarz form of the veilbein \eqref{gss} parameterised as in \eqref{gssexplicit} and integrating the DFT action \eqref{SDFT} along the internal twisted double torus gives an action of the form of (the electric bosonic sector of) gauged half-maximal supergravity   \cite{Arg,geiss}
\bea \label{effective}
S_{d}&=& \frac{1}{2\kappa_d^2}\int d^dx\sqrt{g_d}e^{-2\varphi}
\left[
     {\cal R}+4\partial_\mu\varphi\partial^\mu\varphi-\frac1{12}H_{\mu\nu\rho}
     H^{\mu\nu\rho}\right.  \nn   \\
    && \ \ \ \ \ \  \ \ \ \ \ \  \ \ \ \ \ \ \ \ \ \ \ \ \ \ +\frac 18 D_\mu {\cal H}_{AB}
D^\mu {\cal H}^{AB} -\frac 18{\cal H}^{{}}_{AB}{ F}^{A\mu\nu}
{F}^B_{\mu\nu}\nn\\
&&\left.\ \ \ \ \ \ -\frac 1{12}f_{AB}{}^Cf_{DE}{}^F \left( {\cal H}^{AD}{\cal H}^{BE}{\cal
H}_{CF} -
3
\, {\cal H}^{AD} \eta^{BE} \eta_{CF} + 2 \, \eta^{AD} \eta^{BE} \eta_{CF}
\right)
\right]  .
\eea
Here we have included the kinetic term of the dilaton, ${\cal R}$ is the $d-$dimensional Ricci scalar, $f_{AB}{}^C$ are the constant fluxes, generated by the ``twist" $E_A(y,\tilde y)$ as in \eqref{FABC}, which gauge a subgroup of the global $O(k,k)$ symmetry such that  
\bea \label{covariant}
 D_\mu {\cal H}_{AB}&=&\partial_\mu {\cal H}_{AB}+
\frac{1}{2}f^C{}_{DA} A^D_\mu \nn 
     {\cal H}^{{}}_{CB} + \frac{1}{2}f^C{}_{DB} A^D_\mu{\cal H}^{{}}_{AC} \ , \\
      F^A{}_{\mu\nu} &=& \partial_{[\mu} A_{\nu]}{}^A-\frac{1}{2} f^A{}_{BC} A_{\mu}{}^BA_{\nu}{}^C \ , \\
H_{\mu\nu\rho} &=& 3(\partial_{[\mu} B_{\nu\rho]} -A_{[\mu}^A\partial_{\nu}A_{\rho]A}-f_{ABC}A_{[\mu}^AA_\nu^BA_{\rho]}^C )\nn \ . 
\eea

\section{Effective action from DFT}
\label{effectiveaction}

In this section we show that the generalized Scherk-Schwarz reduction discussed above gives the effective theory of the bosonic string compactified on a torus in the vecinity of a point in moduli space where there is symmetry enhancement. 
We will proceed in two steps. In the first one we do a reduction 
on an ordinary double torus ($i.e.$ no twist) of dimension $2k$, to identify the massless vector and scalar fields of the reduced theory with the corresponding string states of section \ref{sec:massless}. In the second step we show how to incorporate the extra massless states arising at the enhancement point in the DFT description. 

\subsection{Torus reduction}
\label{sec:torus}

As a first step we  consider the generalized Scherk-Schwarz reduction of the previous section on an ordinary double torus ($i.e.$ no twist). The structure constants are zero, and therefore we get an ungauged action with $2k$ abelian vectors  $A_{\mu}{}^A$, and $k^2$ scalars  encoded in ${\cal H}_{AB}$. The $2k$ abelian vectors are those with vertex operators given by \eqref{vertexcartan} and \eqref{vertexcartanR}, corresponding to the $U(1)_L^k \times U(1)_R^k$ symmetry of the torus reduction. The $k^2$ scalars are the fields of the string states given by the vertex operators \eqref{vertexcartanscalars} and are related to the metric and $B$-field on the torus. To get the precise relation, consider an expansion around a given point in moduli space corresponding to a (constant) metric $g_0$ and $B$-field  $B_0$ on the torus. The internal part of the generalized vielbein in the left-right basis ($i.e.$ where
$\eta_{AB}$ has the diagonal form \eqref{Zeta}), given in (\ref{genframeLR}) reads, at first order 
\beq
\begin{pmatrix} {\cal E}_{aL} \\ {\cal E}_{aR} \end{pmatrix} = \frac{1}{\sqrt2} \begin{pmatrix} e_0 -\hat e_0 B_0 & \hat e_0 \\ -e_0 -\hat e_0 B_0 & \hat e_0
\end{pmatrix} +\frac{1}{\sqrt2} \delta \begin{pmatrix} e -\hat e B & \hat e \\ -e -\hat e B & \hat e
\end{pmatrix}\, ,
\eeq
where $e_0$ is a frame for $g_0$.\footnote{\label{foot:ebare}One can actually use a different frame for the left and the right movers, $e_0$ and $\bar e_0$. We will need to use this freedom in section \ref{sec:T2su3gen}.} Performing this expansion and accommodating the terms such that it has the form of a Scherk-Schwarz reduction \eqref{gss}, one gets 
\bea
\begin{pmatrix} {\cal E}_{aL} \\ {\cal E}_{aR} \end{pmatrix}
=  \Phi(x) \  E(y,\tilde y) \, ,\nn
\eea
with
\bea
\Phi(x)=
\begin{pmatrix} 1+ \frac12  (\delta \hat e \, e_0^t+ \delta e \,\hat e_0^t - \hat e_0 \,\delta B\, \hat e_0^t) &
-\frac12 \hat e_0 \, (\delta g - \delta B)\, \hat e_0^t \\
-\frac12 \hat e_0 \, (\delta g + \delta B) \, \hat e_0^t & 1+ \frac12  (\delta e \, \hat e_0^t+ \delta \hat e \, e_0^t + \hat e_0 \, \delta B \, \hat e_0^t)  \end{pmatrix}\nn
\eea
and
\bea
E(y,\tilde y)=\frac{1}{\sqrt2} \begin{pmatrix} e_0 -\hat e_0 B_0 & \hat e_0 \\ -e_0 -\hat e_0 B_0 & \hat e_0 \end{pmatrix}\, ,
\eea
where actually $E$ is constant  here (independent of $y, \tilde y$). We see that $\Phi$ is an element of 
$SO^+(k,k,\mathbb R)$, the component of $O(k,k,\mathbb R)$ connected to the identity.
Inserting this into \eqref{HAB} we get, up to first order, 
\beq \label{HM}
{\cal H}^{AB}=\begin{pmatrix} 1 &
M \\
M^t & 1  \end{pmatrix}\, ,
\eeq
where we have defined
\beq \label{MgB}
M=- \hat e_0 \, (\delta g - \delta B)\, \hat{\bar e}_0^t
\eeq
and 
\beq \label{MtgB}
M^t=- \hat{\bar e}_0 \, (\delta g + \delta B) \, \hat e_0^t\, .
\eeq
Here we are using two different vielbeins, one for the left and another one for the right sectors, $\hat e_0$ and $\hat{\bar e}_0$, giving rise to the same inverse metric $g_0^{-1}$ (see comments below Eq. \eqref{genframeLR}). The scalar fields of the  reduced theory are encoded in the $k \times k$ matrix $M$, which is in the off-diagonal part of ${\cal H}$ (in the left-right basis) and thus has one left-moving and one right-moving index. Writing the indices explicitly, we have 
\begin{equation} \label{Mab}
M_{a b}= - \hat e_{0a}{}^{m} \hat{\bar e}_{o b}{}^{n}\left(\delta g_{mn}-\delta B_{mn}\right).
\end{equation}

Plugging  \eqref{HM}  and taking $f_{AB}{}^C=0$ in \eqref{effective} we get 
\begin{eqnarray}
\label{actionST}
S &=& \frac{1}{2\kappa_d^2} \int d^dx \sqrt{-g} e^{-2\varphi} \left[\mathcal R + 4 \partial^\mu\varphi \partial_\mu\varphi - \frac{1}{12} H_{\mu\nu\rho} H^{\mu\nu\rho} + \frac14\partial_\mu M_{a b} \partial^\mu M^{a b} \right. \nonumber \\
&& \qquad \qquad \quad \left. - \frac{1}{8} F^{a\mu\nu} F_{a\mu\nu} - \frac{1}{8} \bar F^{ a\mu\nu} \bar F_{ a\mu\nu}- \frac{1}{4} M_{a b} F^{a\mu\nu} \bar F^{ b}{}_{\mu\nu}  \right],
\end{eqnarray}
where $F^a$ and $\bar F^{ a}$ are the abelian field strengths for the $U(1)^k_L$ and $U(1)_R^k$ groups, respectively. 
This is precisely the effective action derived from toroidally compactified string theory \cite{polchi}. 

\subsection{Symmetry enhancement}
\label{sec:enhancementDFT}

Here we discuss how to incorporate the symmetry enhancement  arising at special points in moduli space that we discussed in section \ref{sec:massless}. The case of the circle at the self-dual radius, where the $U(1)_L\times U(1)_R$ gauge symmetry is enhanced to $SU(2)_L\times SU(2)_R$, was worked out in \cite{uscircle}. Now we generalize to $T^k$, and an enhancement to a generic group $G \times G$ of dimension $n+n$ and rank $k+k$ (as discussed in section \ref{sec:massless}, the generalization to two different groups with the same rank is straightforward).  The index $a=1,...,n$ runs over the adjoint of $G$, while $A, B=1,...,2n$ and $M,N,=1,\dots ,2n$ run over the adjoint of $G\times G$.

We  incorporate the symmetry enhancement into the DFT formalism through a generalized Scherk-Schwarz reduction, as discussed in section \ref{sec:SS}. To construct the ingredients that go into this recipe, we shall consider the following points.
\begin{enumerate}
	\item The compactification torus will be identified with the maximal torus of the enhanced symmetry group $G$. The $O(k,k,{\mathbb R})$ covariance of the $T^k$ reduction  will be promoted to $O(n,n,{\mathbb R})$, where $n$ is the dimension of $G$.\footnote{Note that the space itself is not extended further than the double torus of dimension $2k$. The derivative in \eqref{GL}  along ``internal directions" has only non-zero components along the $2k$ Cartan directions of the $2n$-dimensional tangent space.} 
	 This has the right dimension to accommodate:
	
	-  the $2n$ massless vector fields $A_{\mu}^a, \bar A_{\mu}^a$ (see \eqref{vertexcartan},\eqref{vertexroots},\eqref{vertexcartanR},\eqref{vertexrootsright}) in the fundamental representation.   
	
	- the $n^2$ scalars $M^{ab}$  (see \eqref{vertexcartanscalars},\eqref{vertexGoldstoneleft},\eqref{vertexGoldstoneright},\eqref{vertexscalars}).
	

	\item We shall assume the ansatz for the generalized frame to have the Scherk-Schwarz form (\ref{gss}), with $A$ in (\ref{gssexplicit})  standing for the $2n$ massless vectors $A_\mu^A$, and $\Phi_A{}^B$ containing the $n^2$ scalars.

	\item The matrix of scalar fields $\Phi_A{}^{B}$ will be assumed to parameterise  $\frac{SO^+(n,n,\mathbb R)}{SO(n,\mathbb R) \times SO(n,\mathbb R)}$ such that  $\mathcal H^{AB}$ in \eqref{HAB} is a symmetric  element of $O(n,n)$ close to the identity, and thus of the form \eqref{HM} in the left-right basis.  The scalar fields are therefore accommodated in the $n\times n$ matrix.
	
		\item The internal piece of the vielbein $E_A{}^{M}$ will be promoted as well to an element in  $O(n,n,\mathbb R)$ and we shall assume that the generalized metric for the background\footnote{By ``the generalized metric for the background" we mean the metric ${\cal G}^{MN}\equiv \delta^{AB}E_A{}^M E_B{}^M$. }, when restricted to the Cartan sector, reduces to the original one, determined by the Cartan metric of the enhanced symmetry group as given by \eqref{metricenhancement}. 
		
	\item The fluxes computed from $E_A{}^{M}$ shall be in diagonal form, {\em i.e.} there is no mixing between barred and unbarred indices, and they reproduce the structure constants of the left and right enhanced gauge groups. A dependence on the internal coordinates is therefore mandatory, but we shall restrict to depence only on the Cartan subsector, namely on the torus coordinates and their duals, $E_A{}^{M}=E_A{}^{M}(y,\tilde y)$. 
	
	\end{enumerate}

Under these considerations, the Scherk-Schwarz reduced DFT action \eqref{effective} reads
\begin{eqnarray}
\label{actionST2}
S &=& \frac{1}{2k_d^2}\int d^dx \sqrt{-g} e^{-2\varphi} \left[\mathcal R + 4 \partial^\mu\varphi \partial_\mu\varphi - \frac{1}{12} H_{\mu\nu\rho} H^{\mu\nu\rho} + \frac14D_\mu M_{ab} D^\mu M^{a b} \right. \nonumber \\
&& \qquad \qquad \qquad \qquad  \quad  - \frac{1}{8} F^{a\mu\nu} F_{a\mu\nu} - \frac{1}{8} \bar F^{ a\mu\nu} \bar F_{ a\mu\nu} 
 - \frac{1}{4} M_{a b} F^{a\mu\nu} \bar F^{ b}{}_{\mu\nu} \\
 &&\qquad \qquad \qquad \qquad  \quad \left. -\frac{1}{12} f_{abc} \bar f_{ a' b' c'} M^{a  a'}M^{b  b'}M^{c  c'} + \frac{1}{4} f_{abc}f^{abc} + \frac{1}{4} \bar f_{abc}\bar f^{abc} \right], \nn 
\end{eqnarray}
where 
\bea \label{DM}
D_\mu M_{a a'} &=& \partial_\mu M_{a a'} + \frac{1}{2} f^{c}{}_{da}A_\mu{}^{d} M_{c a'} + \frac{1}{2} \bar f^{ c'}{}_{ d'  b'} \bar A_\mu{}^{ d'} M_{a c'},  \\
F^a{}_{\mu\nu} &=& \partial_{[\mu} A_{\nu]}{}^a-\frac{1}{2} f^a{}_{bc} A_{\mu}{}^bA_{\nu}{}^c, \nn \\
\bar F^{ a}{}_{\mu\nu} &=& \partial_{[\mu} \bar A_{\nu]}{}^{ a}-\frac{1}{2} \bar f^{ a}{}_{ b c} \bar A_{\mu}{}^{ b} \bar A_{\nu}{}^{ c}, \nn \\
H_{\mu\nu\rho} &=& 3(\partial_{[\mu} B_{\nu\rho]} - A^a{}_{[\mu}\partial_\nu A_{\rho]a}-f_{abc} A^{a}{}_{[\mu}A^b_\nu A^c_{\rho]}-  \bar A^{ a}{}_{[\mu}\partial_\nu \bar A_{\rho] a}-\bar f_{ a b c} \bar A^{ a}{}_{[\mu} \bar A^{ b}_\nu \bar A^{ c}_{\rho]}), \nn 
\eea
Up to the cosmological constant term $\Lambda=\frac{1}{4} f_{abc}f^{abc} + \frac{1}{4} \bar f_{abc}\bar f^{abc}$,
this is the effective action obtained from the three point functions in string theory \cite{ms, uscircle, amn}. The requirement of conformal invariance of the sigma model action (\ref{action}) determines the vanishing of the cosmological constant, and then as argued in \cite{uscircle,hass,dh},  in order to reproduce the string theory results it is necessary to  add 
the $O(D,D)$ covariant  term $-e^{-2d}\Lambda$ to the DFT lagrangian (\ref{SDFT}).

We will show how this action reproduces the right patterns of symmetry breaking when moving away from a point of maximal enhancement, $i.e.$ when giving vacuum expectation values to the Cartan subsector of the matrix of scalars $M$. But before doing that, we pause for a second to expand on the choice of internal vielbein $E_A{}^M (y,\tilde y)$ (or twist) realizing the enhancement of the gauge algebra.     

\subsubsection{Internal double space}
\label{sec:internal}
It was shown in \cite{GM} that the result of a generalized Scherk-Schwarz compactification of DFT is effectively equivalent to  gauging the theory and parameterising the generalized fields in terms of the degrees of freedom of the lower dimensional theory. In this sense, one can simply choose the gaugings in \eqref{actionST2} as the structure constants of the enhanced symmetry algebra. However, the question whether there exists an   
explicit realization of the twists $E_A{}^M$ that gives rise to the enhanced gauge algebra under the generalized diffeomorphisms (\ref{FABC}) is a highly non-trivial one. In this section we discuss this issue and give an explicit realization of the algebra.


A twist for the  ${\mathfrak{su}}(2)^k_L \times {\mathfrak{su}}(2)^k_R$  algebra can be easily found as a straightforward generalization of the one found in \cite{uscircle} for  the circle compactification. Namely, the generalized tangent space is extended so that it transforms in the fundamental representation of $O(d+n, d+n)$, and one can then think of the extra massless vectors with non-trivial momentum and winding as coming from a {\it{metric}} and a {\it B}-field with a leg along these extra $n$ dimensions. The fields in this fictitious manifold depend on a double set of coordinates: $y^m, \tilde y_m, m=1, \dots , k$ dual to the components of  momentum and  winding along the compact directions, respectively. In the left-right basis  it reads
 \bea \label{ESU2k}
E_A{}^M(y_L^m,  y_R^m)&=& \frac{i}{\sqrt2} {\rm diag}\left( J^{ \overline1}, J^{\underline1}, \dots ,  J^{ \overline k}, J^{\underline k},2I_k,-{\overline J}^{\overline1}, -\overline { J}^{\underline 1}, \dots, 
-\overline {J}^{\overline k},- \overline { J}^{\underline  k},-2I_k\right)\, , \  \ \ \  \ \nn\\
 \label{su2vielb}
\eea
where $ J^{\overline m}=J^{\underline m *}=e^{i\sqrt2\alpha_my^m_L} , \overline J^{\overline m}=\overline J^{\underline m *}=e^{i\sqrt2\alpha_my^m_R}$,  are the $SU(2)$ CFT ladder current operators, $i.e.$ there are $k$ pairs of raising and lowering  currents and a $k$ dimensional identity  in the left moving block as well as in the right moving one. This {\it generalized vielbein}
realizes, under the C-bracket of generalized diffeomorphisms
(4.4), $k$ copies of ${\mathfrak {su}}(2)_L$ and $k$ copies of ${\mathfrak {su}}(2)_R$.
To derive this, we use  that the derivative
in (4.4)  has non-trivial
components
\bea
\partial_P=(0,\ 0, \dots , \ 0,\ 0, \ \partial_{y^1_L}, \dots , \partial_{y^k_L},0,\ 0,\  \dots , \ 0,\ 0, \partial_{y^1_R}, \dots,\partial_{y^k_R}),\
\label{gender}
\eea
and  indices are raised and lowered with
\bea \label{eta3}
\eta^{PQ}=\begin{pmatrix}\kappa^{pq}&0\\
0&-\kappa^{pq}
 \end{pmatrix}\, ,
\eea
$\kappa^{pq}$ being the Killing metric of $SU(2)^k$. 

Note that the non-trivial commutator between the raising and lowering directions is obtained from the last term in (\ref{GL}) (or the third term in (\ref{deform})), which is precisely the term that makes the C-bracket differ from the Lie bracket. It is quite remarkable that the C-bracket, which encodes the gauge symmetries of the usual massless sector of the bosonic string theory, $i.e.$ of the states with $N=\bar N=1$, gives rise also to the algebra of the extra massless string modes with momentum and winding, which have $N\ne \bar N$. 
The algebra satisfies
 the quadratic constraints \eqref{quadrconstr}, and therefore the generalized
 Scherk-Schwarz reduction based on this frame is consistent.
Moreover, 
the generalized metric defined from (\ref{su2vielb})  is the identity matrix, and then it reproduces, in particular, the generalized metric of the $k$-torus with diagonal metric and vanishing B-field in the Cartan sector. 

We now generalize this construction for  generic  groups, following the procedure implemented in \cite{uscircle}. 
We start with the generalized vielbein (\ref{genframe})  
\begin{eqnarray}
E_A&=&\left(
\begin{array}{cc}
	 e^a_{\,m} & 0 \\
	-\hat e_a^{\,l} B_{lm} &  \hat e_a^{\,m} 
\end{array}
\right)
\left(
\begin{array}{c}
	 dy^m \\
	\partial_{y^m} 
\end{array}
\right)\, ,
\end{eqnarray}
where $e, \hat e$ and $B$ are the vielbein, inverse vielbein and B-field on the torus at the point of enhancement. Then we  identify $\partial_{y^m}\leftrightarrow d\tilde y_m$, rotate  to the right-left basis on the space-time indices and bring the generalized vielbein to a block-diagonal form  rotating the flat indices, which leads to
\begin{eqnarray}
E_{LR}
&=&\sqrt{2} \left(
\begin{array}{cc}
	 e & 0 \\
	0 & - e
\end{array}
\right)
\left(
\begin{array}{c}
	 dy_L \\
	dy_R
\end{array}\right)\, ,
\end{eqnarray}
where 
\bea
dy_L^m=\frac12 g^{mn}[(g-B)_{np}dy^p+d\tilde y_n]\, , \qquad  d y^m_R=\frac12 g^{mn}[(g+B)_{np}dy^p-d\tilde y_n]\, .
\eea
 Finally, we extend this $2k\times 2k$ matrix so that it becomes an element of $O(n,n)$, where $2n$ is the dimension of $G\times G$. Following the construction implemented above for the $SU(2)^k_L\times SU(2)^k_R$ case, we can  incorporate the CFT ladder currents as\footnote{ We have rearranged the   rows and columns in (\ref{genviel}) so that the extended generalized vielbein looks like (\ref{su2vielb}) and the multiplicative  $i$ factor was added as explained in footnote 11.}
\begin{eqnarray}
E_{LR}(y_L, y_R)
&=&\frac i{\sqrt{2}}
\left(
\begin{array}{cccc}
{\cal J} & 0 & 0 & 0 \\
	 0 & 2 e & 0 &0 \\
  0	& 0 &  - \bar {\cal J}& 0 \\
  0 &0 & 0 &-2e
\end{array}
\right)\, ,\label{genviel}
\end{eqnarray}
where ${\cal J}, \bar{\cal J}$ are $(n-k) \times (n-k)$ diagonal blocks, with $n-k$ the number of  roots of the left and right Lie algebras, with elements 
\bea
{\cal J}_a{}^{i}=\delta_a{}^i \tilde J^{\alpha_i}(y_L) \, , \qquad
\bar {\cal J}_a{}^{i}=\delta_a{}^i \bar {\tilde J}^{\alpha_i}(y_R) \, ,
\end{eqnarray}
and $\tilde J^{{\alpha_i}}(y^1_L,...,y^k_L)=e^{i \sqrt2\alpha_i{}^m  y_L^m}, \bar {\tilde J}^{\alpha_i}(y^1_R,...,y^k_R)=e^{i \sqrt2\alpha_i{}^m  y_R^m}$ are the left- and right-moving raising and lowering currents associated to the $\alpha_i$ root, up to the cocycle factors (see (\ref{vertexroots})). Note that the $2n\times 2n$ matrix 
(\ref{genviel}) depends only on the coordinates associated to the Cartan directions of the algebra. 

We can now calculate the structure constants replacing this generalized frame in (\ref{FABC}), namely
\begin{equation}
f_{ A BC} = 3E_{[A}{}^M \partial_M E_{B}{}^N
E_{C]}{}^
P\eta_{NP} \, .
\end{equation}
Since the currents satisfy $\partial_{y_L^m} \tilde J^{\alpha_i} =i \sqrt2 \alpha_{i}^m\tilde J^{\alpha_i}$ (and similarly for the right-moving  sector), in this way we get
 all the structure constants of G which involve one Cartan generator.

The remaining ones can be obtained in this scheme from the following deformation of the generalized Lie derivative
\bea
({\cal \tilde L}_{E_A}E_B)^M&=& ({\cal L}_{E_A}E_B)^M + \Omega_{AB}{}^CE_C{}^{M} \nn \\
&=&E_{A}{}^N\partial_NE_{B}{}^M- E_B{}^N \partial_N E_A{}^M+ \partial^ME_{A}{}^P E_B{}^Q \eta_{PQ}
+\Omega_{AB}{}^CE_C{}^{M}\, ,\ \ \ \ \ \label{deform}
\eea
where $\Omega_{ABC}$ vanishes if one or more indices correspond to Cartan generators and if $A,B,C$ are associated with roots, say $\alpha,\beta,\gamma$, respectively,
 \bea
\Omega_{ABC}=\left\{\begin{matrix} (-1)^{\alpha * \beta}\;\delta_{\alpha+\beta+\gamma} &\ \ {\rm if \ two \ roots \ are \ positive,}\\
-(-1)^{\alpha * \beta}\;\delta_{\alpha+\beta+\gamma} &\ \ {\rm if \ two \ roots \ are \ negative.} \\
  \end{matrix}\right.\,\nn
 \eea 
This deformation accounts for the cocycle factors that were excluded from  the CFT current operators in (\ref{genviel}) but, as discussed in section 2, they  are necessary  in order to compensate for the minus sign in the OPE $ J^\alpha(z) J^\beta(w)$ when exchanging the two currents and their insertion points $z \leftrightarrow w$ (see Appendix B for more details). It was conjectured in [18] that such factors would also appear in the gauge and duality transformations of double field theory, and actually, they can be included without spoiling
 the local  covariance of the theory. Indeed,  the cocycle tensor $\Omega_{ABC}$ satisfies the consistency constraints of gauged DFT, namely \cite{hk,GM}
\bea
\Omega_{ABC}=\Omega_{[ABC]}\, , \qquad \Omega_{[AB}{}^D\Omega_{C]D}{}^E=0\, , \qquad \Omega_{ABC}\partial^C\cdots =0\, , \label{constraints}
\eea
and it  breaks the $O(n,n)$ global covariance to $O(k,k)$.
Then, all
 the structure constants of $G$ can be obtained  from (\ref{deform}) using  the  expression (\ref{genviel}) for the generalized vielbein.

To see how this works for $SU(3)$, it is convenient to recall the non-vanishing structure constants in the Cartan-Weyl basis
\begin{eqnarray}
&& f_{1\overline 1\underline 1}=f_{1\overline 3\underline 3}=\sqrt\frac32\, , \qquad  f_{2\overline 1\underline 1}=-f_{2\overline 3 \underline 3}=-\frac1{\sqrt2}\, ,\qquad f_{2\overline 2\underline 2}= \sqrt2\, ,\qquad  f_{\overline 1\overline 2\underline 3} =-f_{\underline 1\underline 2\overline 3}=1. \nn
\end{eqnarray}
 These can be obtained from (\ref{deform}) using the  $O(8,8)$ matrix (\ref{genviel})
where
\bea
J^{\overline1}&=&J^{\underline1*}=e^{- i(2y_L^1-y_L^2)}\ , \ \ \ \ \  J^{\overline2}=J^{\underline2*}=e^{- i(y^1_L-2y_L^2)}\ , \ \ \ \ \  J^{\overline3}=J^{\underline3*}=
e^{- i(y^1_L+y_L^2)}\, ,\nn
\eea
and similarly for the right sector, where $y_L^1, y_L^2, y_R^1, y_R^2$ are the coordinates associated to the Cartan directions, and the only non-vanishing components of the cocycle tensor are
\bea
\Omega_{\overline1 \overline2 \underline3}= 1\, , \quad \Omega_{\underline1 \underline2 \overline3}=-1\, .
\eea

Note that the generalized vielbeins (\ref{su2vielb}) and (\ref{genviel}) are eigenvectors of  the operator $\partial_M\partial^M$. Indeed in the LR basis, they verify
\bea
&&-\frac14(\partial_{y^m_L}^2-\partial_{y_R^m}^2)E_{LR}=\nn\\
&& \ \ \ \ \ \ \  \ \ \ =\frac i{2\sqrt{2}}{\rm diag}(\alpha_1^2 J^{\alpha_1} \ ,  \dots , \alpha_{n-k}^2 J^{\alpha_{n-k}}, 0,  \dots  , 
0, \alpha_1^2\bar J^{\alpha_1},   \dots  ,  \alpha_{n-k}^2 \bar J^{\alpha_{n-k}},0, \dots  , 0) \, .\nn
\eea
Interestingly,  this can be written as
\bea
-\frac14(\partial_{y^m_L}^2-\partial_{y_R^m}^2)E_A{}^M=
(N-\bar N)E_A{}^M\, ,
\eea
where  the eigenvalues are $ \frac{\alpha_i^2 }2\,  (-\frac{\alpha^2_i}2)$   in the left-moving (right-moving) sector corresponding to the ladder currents,  and zero in the Cartan sector. This means that the generalized vielbein satisfies a modified version of the weak constraint, holding even when $N\ne \bar N$, which looks like the operator form of the level matching condition  that is necessary to account for the extra massless fields arising at the enhancement point.
The frames  \eqref{ESU2k} and (\ref{genviel}) depend on $y, \tilde y$ and thus they
violate the strong constraint. However, the algebra satisfies
 the quadratic constraints \eqref{quadrconstr}, and therefore the generalized
 Scherk-Schwarz reduction based on this frame is consistent, and  now we see it is also consistent with the level matching condition to be satisfied by the string states. 

We have presented the formalism to obtain  the structure constants in a constructive way. However, once we know the answer, it is instructive to look at it from the DFT point of view. From this perspective, the deformation of the C-bracket given in (\ref{deform}) requires a modification of (\ref{SDFT}), so that  the DFT action is  invariant under the deformed transformations. This would give the action (\ref{actionST2}) with the gaugings  given by $f_{ABC}=\Omega_{ABC}$. Then, a generalized Scherk-Schwarz reduction as described above gives the gaugings containing the Cartan directions and  completes the set of structure constants of the enhanced symmetry group.

\subsubsection{Vielbein \`a la WZW on group manifolds}
\label{sec:vielbeingroupmanifold}

An alternative construction of the generalized internal vielbein can be obtained from the formulation of DFT on group manifolds \cite{DFTgroup,dh} if the fifth condition in section \ref{sec:enhancementDFT} is relaxed and one allows the vielbein to  depend
on coordinates beyond the torus ones and their duals. In this framework, the vielbein depends on the $n$ coordinates of the group manifold corresponding to one of the factors of the enhanced symmetry group. Although this gives a {\it geometric} frame, in the sense that it does not depend on the dual coordinates and then it obeys the strong constraint, it is a natural construction for the WZW model with gauge group $G$ at level 1, which is the  CFT  describing the propagation of the bosonic string on the  corresponding  group manifold.

In this formulation, the vielbein is of the form (\ref{genframeLR}), generalized as in (\ref{HM}) ($i.e.$ we use a different vielbein for the left and the right sectors, $e$, $\bar e$, giving rise to the same metric (see footnote \ref{foot:ebare})), namely \cite{waldramspheres}
\beq \label{genframeLRebare}
\begin{pmatrix} E_{L} \\ E_{R} \end{pmatrix}=\frac{1}{\sqrt{2}}\begin{pmatrix} e -\hat e B & \hat{ e} \\ -\bar e -\hat{\bar e} B & \hat{\bar e}
\end{pmatrix} \ .
\eeq
The frames $e$, $\bar e$ are obtained from the left- and right-invariant Maurer-Cartan forms, taking values on the associated Lie algebra
\begin{subequations}
\begin{equation}
\label{form1}
-i\gamma(y,z)^{-1}\partial_m\gamma(y,z) = e_{m}{}^{a} t_a,
\end{equation}
\begin{equation}
\label{form2}
-i\partial_m\gamma(y,z) \gamma(y,z)^{-1}= \bar e_{m}{}^{\bar a} t_{\bar a},
\end{equation}
\end{subequations}
where we have distinguished between the coordinates $y$ associated to the Cartan generators, and the coordinates $z$ associated to the ladder generators\footnote{Note that the coordinates in \eqref{form1} and \eqref{form2} are the same. There is no dependence on left and right coordinates separately as in the previous section. On the contrary, one needs to introduce the $(n-k)$ coordinates $z$ associated to the roots, although these coordinates have  no natural interpretation from the toroidal geometry.}. Here $\{t_a\}$ is an orthonormal basis for the Lie algebra associated with the gauge group, and it can be freely taken as the standard basis with the inner product defined by the Killing form $\kappa_{ab}$. If $B_{mn}$ is taken such that its field strength in flat indices is given by 
\begin{equation}
H_{abc}=f_{abc},
\qquad
H_{\bar a\bar b\bar c}=f_{\bar a\bar b\bar c},
\end{equation}
the vielbein \eqref{genframeLRebare} gives the desired  $\mathfrak{g} \times \mathfrak{g}$ algebra  \cite{waldramspheres,DFTgroup,hass,dh}. 

We now show that this vielbein also satisfies the assumption (4), namely the generalized metric when restricted to the Cartan subsector reduces to that of the torus. This can be done by choosing a properly factorized parameterisation $\gamma(y,z)$. Let
\begin{equation}
\gamma(y,z)=\Gamma(z)\exp{\left(iy^rh_r\right)},
\end{equation}
where $\{h_r\}$ are the Cartan generators in the Chevalley basis. After replacing it into (\ref{form1}) we obtain
\bea
e_m{}^a t_a&=&-i\exp{\left(-iy^rh_r\right)}\Gamma(z)^{-1}\partial_m\Gamma(z)\exp{\left(iy^rh_r\right)} =-i\exp{\left(-iy^rh_r\right)}\partial_m\exp{\left(iy^rh_r\right)} \nn \\
&=& h_m = \alpha_m{}^{a}t_a,
\eea
where $\alpha_m{}^{a}$ are the simple roots of the Lie algebra. Condition (4) follows from the fact that $\alpha_m{}^{a}\kappa_{ab}\alpha_n{}^{b}=A_{mn}$ is the Cartan matrix. The same can be done for the right sector.
%

Notice that this construction gives $\mathfrak{g}_L=\mathfrak{g}_R$, and there is no straightforward generalization of this construction that yields  different groups\footnote{Different scalings for the structure constants on the left and on the right sector can be obtained by allowing dependence on one dual coordinate \cite{wso8}. However, as pointed out in \cite{DFTgroup}, different scalings correponds to having different levels in the Kac-Moody algebra of left and right-movers and thus has no connection to the CFT of the bosonic string on a torus.}.

\section{Effective description around a point of maximal enhancement}
\label{sec:T2su3gen}
In this section we show that the effective action \eqref{actionST2} reproduces the string theory results at the vecinity of a particular point in moduli space where the symmetry enhancement is maximal, $i.e.$ given by $G \times G$, with $G$ a semi-simple and simply laced Lie group of dimension $n$ and rank $k$. We check in particular that it gives the right masses of scalar and vector fields  when moving slightly away from the point of symmetry enhancement.  We do this in general for any $T^k$ and $G \times G$ group, and then inspect more closely the example of $T^2$ at the $\stt$ point, which is the simplest setup  that has already all the non-trivialities of the generic case. All the necessary definitions and conventions of Lie algebras are given in Appendix \ref{app:Liealgebras}.  

Let us first recall the notation. We use $a,b=1,...n$ to denote indices in the adjoint representation, and $p,q=1,..,k$ for the Cartan subsector. We will use the Chevalley basis defined in \eqref{defh} where the triplet $e^{\underline p}, h^p, f^{\overline p}$ of raising, Cartan and lowering generators associated to the simple root $p$ satisfy the standard $\mathfrak{su}(2)$ commutation relations. Additionally, we use an index $\overline u=1,...,\frac12(n-3k)$ to denote the lowering generators associated to the non-simple roots. The index $\overline \imath=1,...,\tfrac12 (n-k)$ labels all negative roots (and thus $\overline \imath=\{\overline p,\overline u\}$). 

The $2n$ vectors that span the enhanced symmetry group $G\times G$ are split into two types:
\begin{enumerate}
\item $2k$ real vectors $A^{p}, \bar A^{p}$ along the $k$ Cartan directions
\item $2 \times \frac12(n-k)$ complex vectors $A^{\overline \imath}, \bar A^{\overline \imath}$  along ladder generators ($A^{\underline i}=(A^{\overline \imath})^\dagger$) which can be further split into simple and non-simple roots $A^{\overline \imath}=\{A^{\overline p}, A^{\overline u}\}$ and  $\bar A^{\overline \imath}=\{\bar A^{\overline p}, \bar A^{\overline u}\}$
\end{enumerate}

The $n^2$ scalars are  split into three types:
\begin{enumerate}
\item $k^2$ real scalars of the form $M^{pp'}$ with both legs along a Cartan direction
\item $k(n-k)$ complex scalars of the form $M^{p\overline \imath}$ with one leg along a Cartan direction
\item $(n-k)^2$ complex scalars of the form $M^{\overline \imath \overline \jmath}$ with no legs along Cartan directions
\end{enumerate}

The vectors in (1) are massless at all points in moduli space and parameterise $g_{\mu p} \pm B_{\mu p}$. The vectors in (2) get masses when moving away from the point of symmetry enhancement, as we will discuss in detail. The scalars in (1) are the ones that remain massless at all points in moduli space. They parameterise the metric and $B$-field on the $k$-torus, or rather its deviation from the value at the enhancement point (which we take to be the origin in moduli space).  
In the neighborhood of an  enhancement point  these scalars acquire a vacuum expectation value $v^{pp'}$. We thus redefine
\beq \label{vevs}
M^{pp'} \to 4 v^{pp'} + M^{pp'}\, ,
\eeq
so that $\left\langle M^{aa'} \right\rangle=0$ for all $a,a'$. The factor of $4$ is introduced in order to compare the results with those coming from string theory. We show now in detail how these vevs break the enhanced symmetry spontaneously.
The covariant derivative of the scalars given in \eqref{DM} becomes, for those with one index along a Cartan direction 
\beq
D_\mu M_{a a'} &=& \partial_\mu M_{a a'} + \frac{1}{2} f^{c}{}_{da}A_\mu{}^{d} M_{c a'} + \frac{1}{2} \bar f^{ c'}{}_{ d'  b'} \bar A_\mu{}^{ d'} M_{a c'}\, .
\eeq
The square of this gives a mass to the vectors $A^{\overline p}$. Similar terms give masses to all vectors associated to ladder generators. We get\footnote{Again, in order to compare with string theory later on, we should take the structure constants in the DFT effective action equal to one half the structure constants in the Chevalley basis. This can be done by rescaling all the generators: $J^a\rightarrow \frac{1}{2} J^a$.}  
\begin{subequations} \label{Mvectors}
\bea
m^2_{A^{\overline p}}& =& v^{(2)}{}^{pp} \ , \quad \quad m^2_{A^{\overline u}} =  (n^t  v^{(2)} n)^{uu}\, ,  \label{MvectorsL}
\\
m^2_{\bar A^{\overline{p}'}} &=& (v^{(2)}{}^t)^{p'p'}  \ , \quad m^2_{\bar A^{\overline{u}'}}=   (n^t  v^{(2)}{}^t n)^{u'u'}  \, , \label{MvectorsR}
\eea
\end{subequations}
where we have defined 
\beq \label{V2}
v^{(2)} = v A^{-1} v^t \, ,
\eeq
and $v$ is the $k\times k$ matrix of vevs, $v^{pp'}$ in the Chevalley basis, while $n$ is the matrix of coefficients $n_p{}^u$  in the linear combination of the root $\alpha_u$  
\beq \label{cpu}
\alpha_u=n_p{}^u \alpha_p \ . 
\eeq

Some of the scalars acquire masses from the potential term $f\bar f MMM$ in the effective action \eqref{actionST2}, which becomes \begin{eqnarray} \label{Vgen}
V_{f \bar fMMM} &=&-\frac{1}{2}\left\{ v^{pp'} \left[ (|M_{\overline p\overline p'}|^2- | M_{\overline p \underline p'}|^2)+  n_{p'}{}^{u'} (|M_{\overline p \overline u'}|^2-  |M_{\overline p \underline u'}|^2 ) \right.\right.\nn\\
&&\left.\left.\qquad + \  n_{p}{}^u (|M_{\overline u \overline p'}|^2-  |M_{\overline u \underline p'}|^2 )  
 +  n_p{}^{u} n_{p'}{}^{u'} (|M_{\overline u\overline u'}|^2- |M_{\overline u\underline u'}|^2) \right] \right\}\nn\\
&&\qquad + \ f_{abc}f_{a'b'c'}M^{aa'}M^{bb'}M^{cc'}\, ,
\end{eqnarray}
where we have used the structure constants in \eqref{eq:simple}, \eqref{eq:nonsimple}. 
We see that only scalars of type (3) (with no legs along Cartan directions) acquire masses from this term, with a mass squared proportional to the vev's. 
Scalars of type (2) remain massless at this level and are the Goldstone bosons of the spontaneous symmetry breaking. As we will see shortly, they acquire masses that are quadratic in the vevs, matching those of the vectors.

We discuss now in more detail the process of spontaneous symmetry breaking. 
By giving arbitrary vevs to all scalars in the Cartan subsector, we see that all the gauge vectors acquire mass and the gauge symmetry is spontaneosly broken to $U(1)_L^k \times U(1)_R^k$. 
Similarly, if $v$ has a row with all zeros, let's say the row $p$, then the corresponding (complex) vector $A^{\underline p}$ remains massless, and there is an $SU(2)$ subgroup of $G_L$ that remains unbroken. 
If the matrix $v$ has a column with all zeros, the  vectors $\bar A^{\underline p}$ of the right-moving enhanced gauge symmetry stay massless and there is an $SU(2)$ subgroup of $G_R$ that stays unbroken\footnote{In the following, whenever not necessary, we do not make the distinction between a positive or a negative root, denoted respectively by $\underline i$ and $\overline \imath$.}. Since the Cartan matrix is non-degenerate, the converse is also true: the only way that a vector $A^{\underline p}$ remains massless is if $v^{pp'}=0$ for all $p'$: 
\bea
v^{p p'}=0 \ \  \forall p' \qquad \Leftrightarrow \qquad m^2_{A^{\underline p}} =0  ,\,\nn \\
v^{p p'}=0 \ \  \forall p \qquad \Leftrightarrow \qquad m^2_{\bar A^{\underline p\prime}} =0 \, .\label{converse}
\eea
For the vectors associated to non-simple roots the situation is more tricky as it depends on which integers $n_p{}^u$ are non-zero. $A^{\underline u}$ remains massless if $v^{pp'}=0$ for all $p$ such that $n_p{}^u\neq 0$ and for all $p'$. Note that we cannot preserve only the vectors corresponding to non-simple roots: if all the vectors corresponding to simple roots are massless, then necessarily $v=0$ and then there is no symmetry breaking at all. This implies that the spontaneus breaking of symmetry always involves at least one $U(1)$ factor, corresponding to the Cartan of the $SU(2)$ associated to the simple root whose vector becomes massive. Thus we cannot go from one point in moduli space of maximal enhancement (given by a semi-simple group) to another of maximal enhancement by a spontaneous breaking of symmetry. We will comment more on this in the next section.          

Regarding the scalars, arbitrary vevs (for scalars of type (1)) give masses to scalars of the type (3), with no legs along Cartan directions. The squared masses are linear in the vevs. Scalars of type (2), with one leg along a Cartan direction, stay massless at this level. However, it is easy to see that they acquire a mass of the same order of the vectors, namely second order in the vevs, coming from expanding ${\cal H}^{AB}$ in \eqref{effective} to second order in $M$, which gives (cf. \eqref{HM} for the expansion at first order)
\beq \label{HM2}
{\cal H}^{AB}=\begin{pmatrix} 1+\tfrac12 MM^t &
M \\
M^t & 1+\tfrac12 M^t M  \end{pmatrix}\, .
\eeq
From the term $ff{\cal H}{\cal H}{\cal H}$ in \eqref{effective} we get a contribution that is quartic in $M$ coming from two factors of ${\cal H}$ expanded at second order and the third factor at zeroth order. This gives masses to the scalars that are precisely those of the vectors, namely we get (including also the linear order contribution) 
\begin{eqnarray}
&&m^2_{M^{\underline p \underline p\prime }}= - 2 v^{pp\prime} + m^2_{A^{\underline p}}+m^2_{\bar A^{\underline p\prime}}\, , \nonumber\\
&&m^2_{M^{\underline p \underline u\prime}}= - 2 v^{pp\prime} n_{p\prime}{}^{u\prime} + m^2_{A^{\underline p}}+m^2_{\bar A^{\underline u\prime}} \, ,\nn\\
&&m^2_{M^{\underline u \underline u\prime}}= - 2 n_p{}^u v^{pp\prime}   n_{p\prime}{}^{u\prime}+ m^2_{A^{\underline u}}+m^2_{\bar A^{\underline u\prime}}\, , \nn \\
&&m^2_{M^{p \underline {i\prime}}}= m^2_{\bar A^{\underline i\prime}} \, .\nonumber 
\label{goldstonemass}
\end{eqnarray}
We thus get that for arbitrary vevs, all vectors and scalars except those along Cartan directions acquire masses, and the symmetry is broken to $U(1)_L^k \times U(1)_R^k$. If $v^{pq}=0$ for a given $p$ and for all $q$, while all other vevs are non-zero, then the remaining symmetry is $(SU(2) \times U(1)^{k-1})_L \times U(1)_R^k$ where the $SU(2)_L$ factor corresponds to the root $p$,   
and the massless scalars are, besides those purely along Cartan directions, those of the form $M^{\underline p q}$. 

\subsection{Comparison with string theory}

Let us compare the vector and scalar masses that we got in the previous section from the double field theory effective action to those of string theory, given by \eqref{MZhZ}. 
We decompose the generalized metric ${\cal H}$ as in \eqref{HMHA} where $E_A{}^M$ is the generalized vielbein at the point of enhancement, that we will call $E_0$, and ${\cal H}_{AB}$ represents the fluctuations from the point, parameterised in terms of the matrix $M$ as in \eqref{HM} (though here we will need the second order term as well, the expression up to second order is given in \eqref{HM2}\footnote{Note that $M$ here is a $k\times k$ matrix spanning along the Cartan directions only, as in section \ref{sec:torus}. }). Inserting this in the mass formula we get 
\beq \label{MZhZE}
m^2=2 ( N + \bar N -2) +  Z^t E^t_{0} \begin{pmatrix} 1+\tfrac12 MM^t &
M \\
M^t & 1+\tfrac12 M^t M  \end{pmatrix}
 E_{0} Z \ .
\eeq
On the other hand, from Eq. \eqref{pE} 
\beq
 E_{0} \, Z =  \begin{pmatrix} p_{aL} \\ p_{aR} \end{pmatrix} \ .
\eeq
We thus get
\bea \label{MM}
m^2&=&2 ( N + \bar N -2) +  \\
&& p^t_{L} (1+\tfrac12 MM^t) p_L +  p^t_{R} (1+\tfrac12 M^tM) p_R+ p^t_{L} M p_R + p^t_{R} M^t p_L \ . \nn 
\eea
For left-moving vectors one has $\bar N=1$ and either $N=1$ and $p_L=p_R=0$, or $N=0$ and $p_{R}=0$, $p_{aL}=\alpha^{a}$ with $\alpha$ a root of the enhanced gauge algebra (which have norm  $p_L^t p_L=2$). The former vectors (Cartans) are massless for any $M$, while the latter have mass
\beq
m_{A^{\alpha}}^2=\tfrac12 \alpha^t MM^t \alpha \ .
\eeq 
Similarly, for the vectors in the right sector, we get
\beq
m_{\bar A^{\alpha}}^2=\tfrac12 \alpha^t M^tM \alpha \ .
\eeq 
Inserting the expression \eqref{MgB} for the matrix $M$ in terms of $\delta g - \delta B\equiv v$
we get
\beq
m_{A^{\alpha}}^2=\frac12 \alpha^t \hat e_0 v g_0^{-1} v^t \hat e^t_0 \alpha = \alpha^t \hat e_0 v A^{-1} v^t \hat e^t_0 \alpha = \alpha^a v^{(2)}_{ab} \alpha^b \ ,
\eeq 
where in the second equality we have used that the metric at the points of maximal enhancement is given by the Cartan matrix, Eq. \eqref{metricenhancement}, and in the third equality the definition of $v^{(2)}$ given in \eqref{V2}. 
This is precisely the mass for the left moving vectors found from DFT, Eq. \eqref{Mvectors}, where for simple roots the contraction of $v^{(2)}$ in an orthonormal basis with a root $\alpha_p$, gives the component $(v^{(2)})^{pp}$ in the Chevalley basis as in \eqref{Mvectors}. For a non-simple root $\alpha_u$ we have to contract additionally with the components of $\alpha_u$ in the basis of simple roots, given by the integers $n$, and we recover again the masses in \eqref{Mvectors}.    

For the scalars, those of type (1) (both legs along Cartan directions) have $N=\bar N=1$, $p_L=p_R=0$, and are massless for any $M$. Scalars of the form (2) (one leg along Cartan) have $N=1$ and $p_L=0, p_R=\alpha$ (or the same exchanging left and right). Their masses are thus exactly those of the vectors corresponding to the same root, namely
\beq
m^2_{M_{p \underline i'}}=m^2_{\bar A^{\underline i'}} \ , \qquad m^2_{M_{\underline i p}}=m^2_{ A^{\underline i}}\, ,
\eeq
in agreement with what we have found from DFT, Eq. \eqref{goldstonemass}, confirming that these are the Goldstone bosons of the spotanteous breaking of symmetry. Finally, scalars of type (3) have $N=\bar N=0$, $p_L=\alpha$, $p_R=\beta$ and their masses are
\beq
m^2_{M_{\alpha \beta}} = \alpha^t M \beta + \beta^t M^t \alpha + \tfrac12 \alpha^t   MM^t \alpha +\tfrac12 \beta^t  M^tM \beta \, .
\eeq
Using again \eqref{MgB} to write this in terms of $v$ we get
\bea \label{Mscalars}
m^2_{M_{\alpha \beta}} & =& -2 \, \alpha^a v_{ab} \beta^b +\alpha^a  v^{(2)}_{ab} \alpha^b + \beta^a  v^{(2)t}_{ab} \beta^b \nn \\
 & =& -2 \, \alpha^a v_{ab} \beta^b + m^2_{A^{\alpha}} + m^2_{\bar A^{\beta}} \ . 
\eea
This precisely matches the masses found from DFT, Eq. \eqref{goldstonemass}, as, for example, taking $\alpha=\alpha_p$, $\beta=\alpha_{p'}$ 
$\alpha^a V_{ab} \beta^b=v^{pp'}$.

We have shown that the masses found from DFT precisely resproduce the string theory masses, up to second order in the vacuum expectation value of the scalars along Cartan directions. We have identified these vacuum expectation values with the fluctuations of the metric and $B$-field away from the point of maximal enhancement, namely
\beq \label{V}
 v= \delta g - \delta B \ .
\eeq

In the next section we will show this matching even more explicitly for the case of  $T^2$ at the $SU(3)_L \times SU(3)_R$ enhancement point. 

\subsection{Example of $T^2$ around $SU(3)_L \times SU(3)_R$ enhancement point}
\label{sec:T2su3}

Let us be more explicit with certain aspects discussed in the previous section by inspecting the effective action for a $T^2$ around the $\stt$ enhancement point. 
The relevant formulas for the  $\mathfrak{su}(3)$ algebra are given in \ref{app:SU3}. Here $a=1,..,8$, $i=1,2$ and the index $u$ has a single value that we call 3. 

The masses for the vectors in the left-moving sector \eqref{MvectorsL} are 
\begin{eqnarray}
m_{A^{\overline 1}}^2&=& \tfrac23 ((v^{11})^2+v^{11}v^{12}+(v^{12})^2)  ,\,\nn \\
m_{A^{\overline 2}}^2&=&\tfrac23 ((v^{21})^2+v^{21}v^{22}+(v^{22})^2)  \, , \\
m_{A^{\overline 3}}^2&=&  \tfrac23 ((v^{11})^2+v^{11}v^{12}+(v^{12})^2+(v^{21})^2+v^{21}v^{22}+(v^{22})^2+ \nn\\
&&2v^{ 1 1}v^{ 2 1}+v^{ 1 1}v^{ 2 2}+v^{ 1 2}v^{ 2 1}+2v^{ 1 2}v^{ 2 2}) \, ,\nn
\label{eq:masasleft}
\end{eqnarray}
where the vevs are given in the Chevalley basis. For the right-moving sector we have 
\begin{eqnarray}
m_{\bar A^{\overline 1}}^2&=& \tfrac23 ((v^{11}))^2+v^{11}v^{21}+(v^{21})^2) \, ,\nn \\
m_{\bar A^{\overline 2}}^2&=&\tfrac23 ((v^{12})^2+v^{12}v^{22}+(v^{22})^2) \, ,  \\
m_{\bar A^{\overline 3}}^2&=& \tfrac23  ((v^{11})^2+v^{11}v^{21}+(v^{21})^2+(v^{12})^2+v^{12}v^{22}+(v^{22})^2+ \nn\\
&&2v^{11}v^{12}+v^{11}v^{22}+v^{21}v^{12}+2v^{21}v^{22})\, .\nn
\label{eq:masasright}
\end{eqnarray}

The potential  \eqref{Vgen} reads
\begin{eqnarray} \label{VSU3}
 V &=&-\frac{1}{2}\left[ v^{11} (|M_{\underline 1\underline 1}|^2- |M_{\underline 1\overline 1}|^2)+v^{12} (|M_{\underline 1\underline 2}|^2- |M_{\underline 1\overline 2}|^2)+ \right.\nonumber \\
&& \left.v^{21} (|M_{\underline 2\underline 1}|^2- |M_{\underline 2\overline 1}|^2)+v^{22} (|M_{\underline 2\underline 2}|^2- |M_{\underline 2\overline 2}|^2)+ \right.\nonumber \\
&& \left.(v^{11}+v^{12}) (|M_{\underline 1\underline 3}|^2 -  |M_{\underline 1\overline 3}|^2)+(v^{11}+v^{21}) (|M_{\underline 3\underline 1}|^2 -  |M_{\underline 3\overline 1}|^2 +\right.  \\
&&\left. (v^{12}+v^{22}) (|M_{\underline 3\underline 2}|^2 - |M_{\underline 3\overline 2}|^2) +(v^{21}+v^{22}) (|M_{\underline 2\underline 3}|^2 - |M_{\underline 2\overline 3}|^2)+\right. \nonumber \\
&& \left.(v^{11}+v^{12}+v^{21}+v^{22}) (|M_{\underline 3\underline 3}|^2 - |M_{\underline 3\overline 3}|^2)\right] + f_{abc} \bar f^{a'b'c'} M^{aa'}M^{bb'}M^{cc'}.\nonumber
\end{eqnarray}

We can see explicitly what we discussed in the previous section. Giving vevs to all $v^{pp'}$ breaks the full $SU(3)_L \times SU(3)_R$ to $U(1)_L^2 \times U(1)_R^2$. 
If $v^{11}=v^{12}=0$, and the others are non-zero, we break to $(SU(2) \times U(1))_L \times (U(1)^2)_R$, where the $SU(2)_L$ that is preserved corresponds to the root 1. The scalars $M_{\underline1 \underline i}$ stay massless at linear order in the vevs, while the other scalars with no roots along the Cartans are massive.  As discussed in the previous section, the scalars $M_{2 \underline i}$ acquire a mass at second order in the vevs, equal to that of the vector $\bar A^{\underline i}$, and are the Goldstone bosons for the symmetry breaking on the right sector. The scalars  $M_{\underline1 \underline i}$ also acquire a second order mass given by the mass of the vectors $\bar A^{\underline i}$.  
Similarly, for $v^{21}=v^{22}=0$ and the others non-zero we preserve the same group, but the unbroken  $SU(2)_L$ corresponds to the second root.
For $v^{11}=v^{21}=0$ we preserve   $(U(1)^2)_L \times (SU(2) \times U(1))_R$, where the $SU(2)_R$  corresponds to the first root, and the massless scalars are $M_{\underline i \underline 1}$. In order to break to $(SU(2) \times U(1))_L \times (SU(2) \times U(1))_R$ we need to set three vevs to zero, and the only non-zero vev is along the Cartans of the two massive vectors ($i.e.$ only $v^{22}\neq 0$ if the two preserved $SU(2)$'s are those of the root 1).  
We also see that we cannot have the third vector massless while the other two massive, as we anticipated in the previous section. We thus cannot break to $SU(2)^2_L \times SU(2)^2_R$.



Let us compare now to the situation in string theory, discussed in detail in section \ref{sec:stt}. The four vevs $v^{pp'}$ correspond to the fluctuations of $g$ and $B$ away from the enhancement point, which in the case of  $T^2$ are parameterised by the two complex moduli $\tau$ and $\rho$, or rather by $\delta \tau=\tau - (-\tfrac12 + i \sqrt{3}), \ \delta \rho=\rho - (-\tfrac12 + i \sqrt{3})$. We can see from Figure 2  that in order to preserve $(SU(2) \times U(1))_L \times (SU(2) \times U(1))_R$ we need to move  either along the curve $|\tau|=|\rho|=1$, or along the vertical line $\tau_1=\rho_1=-\tfrac12$. Both curves are one-dimensional and thus given in terms of a single parameter, or a single vev, in accordance to what we just discussed. Staying away from these lines, but still on the plane $\tau=\rho$, requires two independent parameters, in accordance to the statement that in order to break to 
$U(1)^2_L \times (SU(2) \times U(1))_R$ one needs two vevs. 

We now find the relation between the vevs $v^{pp'}$ and $\delta \tau, \ \delta \rho$ directly by comparing the DFT and the string theory masses for the scalars at linear order in the vevs, and then show that we indeed get equation \eqref{V} relating the vevs to the metric and $B$-field.    
Setting equal  the DFT and string theory masses (that we write on the left and right hand side, respectively) for the following scalars, whose momentum and winding numbers are given in Table \ref{ta:masslessscalarsSU3}, we get 
\begin{eqnarray}
&&M_{\underline 1 \underline 1}:\quad v^{11}= - \sqrt{3} \delta \rho_2+3 \delta \tau_1-\sqrt{3} \delta \tau_2\, ,\nn\\
&&M_{\underline 2 \underline 2}:\quad v^{22}= 2 \sqrt{3} \delta \tau_2-2 \sqrt{3} \delta \rho_2\, ,\nn\\
&&M_{\underline 3 \underline 3}:\quad v^{11}+ v^{12}+ v^{21}+ v^{22}= -2 \sqrt{3} \delta \rho_2-3 \delta \tau_1-\sqrt{3} \delta \tau_2\, ,\nn\\
&&M_{\underline 2 \underline 3}:\quad v^{21}+ v^{22}= -3 \delta \rho_1-\sqrt{3} \delta \rho_2-3 \delta \tau_1+\sqrt{3} \delta \tau_2\nn\, .
\label{eq:}
\end{eqnarray}
And thus we identify, in the Chevalley basis,
\beq \label{Vsu3}
v=\frac43 \begin{pmatrix} 3  \delta \tau _1 - \sqrt{3}(2 \delta \rho _2+  \delta \tau _2) & 3  (\delta \rho _1-  \delta \tau _1)+\sqrt{3} ( \delta \rho _2-  \delta \tau _2) \\
-3  (\delta \rho _1+  \delta \tau _1)+\sqrt{3} ( \delta \rho _2-  \delta \tau _2)   & - 2 \sqrt{3} (  \delta \rho _2-   \delta \tau _2)
\end{pmatrix}\, .
\eeq
We can see from here the different directions of symmetry breaking: the plane $\tau=\rho$ has $\delta \tau=\delta \rho$ and thus $v^{12}=v^{22}=0$. Therefore, on this plane the $SU(2)_R$ corresponding to the root 2 is unbroken. On the vertical line one has additionally $\delta \tau_1=\delta \rho_1=0$, and thus only $v^{11}\neq 0$, and preserving additionally the $SU(2)_L$ symmetry corresponding to the root 2. The curve $|\tau|=1$ has $\delta \tau_1=-\frac{\tau_2}{\tau_1} \delta \tau_2=\sqrt{3} \delta \tau_2$ and thus only $v^{21}\neq 0$, so the $(SU(2) \times U(1))_L \times (SU(2) \times U(1))_R$ unbroken symmetry corresponds to the root 1 on the left and 2 on the right.


We can also check that we recover the masses of the vectors coming from string theory. Replacing these expressions into the mass for the vectors \eqref{eq:masasleft} we get the string theory masses expanded at second order in the fluctuations 
\begin{eqnarray}
&&m^2_{A^1_l}=\frac{4}{3}\left(\delta \rho_1^2-\delta \rho_1 \delta \tau_1-\sqrt{3} \delta \rho_1 \delta \tau_2+\delta \rho_2^2-\sqrt{3} \delta \rho_2 \delta \tau_1+\delta \rho_2 \delta \tau_2+\delta \tau_1^2+\delta \tau_2^2\right)\, ,\nn\\
&& m^2_{A^2_l}=\frac{4}{3}\left(\delta \rho_1^2+2 \delta \rho_1 \delta \tau_1+\delta \rho_2^2-2 \delta \rho_2 \delta \tau_2+\delta \tau_1^2+\delta \tau_2^2\right) \, ,\\ 
&&m^2_{A^1_l}=\frac{4}{3}\left(\delta \rho_1^2-\delta \rho_1 \delta \tau_1+\sqrt{3} \delta \rho_1 \delta \tau_2+\delta \rho_2^2+\sqrt{3} \delta \rho_2 \delta \tau_1+\delta \rho_2 \delta \tau_2+\delta \tau_1^2+\delta \tau_2^2 \right)\, .\nn
\label{eq:}
\end{eqnarray}

We now check that the expression  we obtained for $v$ is in agreement with \eqref{V}. Recall, from \eqref{Mab} we have
\beq
- (e_0)^a{}_{p} M_{ab} (e_0)^b{}_{p'}= - M_{pp'} = \ (\delta g - \delta B)_{pp'}.
\eeq
Taking the expectation value of this, and using \eqref{vevs}
\beq 
M^{pp'} \to 4 v^{pp'} + M^{pp'}\,, \nn
\eeq
we get
\beq 
-4 (e_0)^a{}_{p} v_{ab} (e_0)^b{}_{p'} = \ (\delta g - \delta B)_{pp'}.
\eeq
Up to first order in $\delta \rho$ and $\delta \tau$ one has
\beq 
v_{ab}= \frac2{\sqrt{3}} \left( \begin{array}{cc}
	\delta\rho_2+ \delta\tau_2  &-\delta\rho_1+ \delta\tau_1 \\
	\delta\rho_1+ \delta\tau_1 & \delta\rho_2-\delta\tau_2
\end{array}
\right)
\eeq
\beq \label{finito}
(\delta g - \delta B)_{\rm{Chevalley}}= -4 e^{t}  \frac2{\sqrt{3}} \left( \begin{array}{cc}
	\delta\rho_2+ \delta\tau_2  &-\delta\rho_1+ \delta\tau_1 \\
	\delta\rho_1+ \delta\tau_1 & \delta\rho_2-\delta\tau_2
\end{array}
\right) e =   \eqref{Vsu3}  \, ,
\eeq
where 
\beq
e= \begin{pmatrix} \sqrt{3}/2 & 0 \\ -1/2 & 1 \end{pmatrix}
\eeq
is the vielbein at the $SU(3)_L \times SU(3)_R$ enhancement point.  We have thus verified Eq. \eqref{V}, namely the vacuum expectaction value of the scalar fields with two Cartan indices are nothing but the deviation of metric and $B$-field away from the point of symmetry enhancement.

\section{Effective description for all moduli space}
\label{sec:Td}

Up to this point we have been able to describe the effective behavior of the bosonic string compactified on a torus using DFT strategies at any point in moduli space. For points with no symmetry enhancement, the usual (ungauged) DFT formulation directly gives the answer. Symmetry enhancement, on the other hand, requires an ad-hoc increase of the dimension of the $O(d+k,d+k)$ realization in order to account for the full massless sector in the low energy regime. Spontaneous symmetry breaking provides a suitable process for further exploring the neighborhood of each enhancement point in moduli space.

If symmetry breaking in target space can be integrated beyond the infinitesimal level\footnote{A hint that this can be the case is given by the fact that spontaneous symmetry breaking is related to exactly marginal deformations of the associated WZW model.}, {\em i.e.} vevs can acquire a finite value, it can be realized as a process for describing not only a neighborhood of the enhancement point but an extended ``patch'' in moduli space. A natural question arising is if it is possible for this patch to cover the full moduli space. Or equivalently, if there exists a unique DFT describing the entire moduli space this way.

For the simplest case of the compactification on a circle, the answer is positive. Indeed, the DFT action for the bosonic string compactified on a self-dual circle, which is associated with the unique enhancement point in the corresponding moduli space, gives rise after symmetry breaking to the action for any other circle compactification \cite{uscircle}. However, for a compactification on a higher dimensional torus, if we insist in looking for a unified description in a maximal enhancement point, the answer fails to be positive already for $2$-dimensional tori. As we have seen, maximal enhancement points for a $2$-torus are those corresponding to $SU(3)_L\times SU(3)_R$ and $SU(2)_L^2 \times SU(2)_R^2$, the first group having higher dimension than the second one. But $SU(2)^2$ cannot be realized as a regular subgroup of $SU(3)$, and therefore points with $SU(3)_L\times SU(3)_R$ and $SU(2)_L^2 \times SU(2)_R^2$ cannot be linked by a symmetry breaking, a fact that has been already pointed out in a more general setting in a previous section. Neither $SU(3)_L\times SU(3)_R$  nor $SU(2)_L^2\times SU(2)_R^2$ could describe the full moduli space.

In order to accommodate all maximal enhancement points in a single approach, we shall need additional dimensions. How the new higher dimensional DFT should be implemented and how we need to treat the new ``unphysical'' directions is the subject of the rest of this section.

For simplicity, let us discuss, again, the $2$-dimensional torus. Even though the vectors  that are purely left-moving at one point in moduli space become a mixture of left and right-moving vectors away from that point, left-moving massless vectors never become purely right moving  at some other point. We thus treat the left-moving and right-moving vectors independently in what follows. As discussed, $SU(2)^2$ is not a regular subgroup of $SU(3)$, but both groups can be embedded in the rank $3$ group $SU(2)\times SU(3)$. Let us therefore consider a $3$-dimensional toroidal compactification with $(SU(2)\times SU(3))_L\times (SU(2)\times SU(3))_R$ enhancement. The action obtained through DFT strategies with $O(d'+11,d'+11)$ symmetry is\footnote{Here $d'$ stands for the dimension of the extended space and $11$ is the dimension of $SU(2)\times SU(3)$. We do not use the previous notation for reasons that will become clear in what follows.} \eqref{actionST2} with $a,b,\dots,a',b',\dots$ running in the adjoint representation of $SU(2)\times SU(3)$. The structure constants are explicitly given in Appendix A.

If we set all Cartan scalar vevs to zero except $v^{11}$, $SU(2)\times SU(3)$ spontaneously breaks into $U(1)\times SU(3)$. Indeed, since we have $v^{2p'}=v^{3p'}=0$ for any $p'$, from \eqref{converse} follows that $A^{\underline{2}}$ and $A^{\underline 3}$ remain massless and the same is true for $\overline A^{\underline{2}}, \overline A^{\underline{3}}$ since $v^{p2}=v^{p3}=0$ for all $p$. The vector $A^{\underline{4}}$ also remains massless since $n_p{}^4\ne 0$ for $p=2,3$. On the other hand, vectors indexed with a Cartan entry never get mass. Therefore, only vectors with a leg in a root of $SU(2)$, namely, $A^{\underline{1}}$, $A^{\overline{1}}$, $\overline A^{\underline{1}}$ and $\overline A^{\overline{1}}$, acquire mass through (5.2). In turn, some scalars also acquire mass. These are those scalars with at least a leg in a root of $SU(2)$. More explicitly, $M^{\underline p \underline p'}$, $M^{\underline p \overline p'}$, $M^{\overline p \underline p'}$ and $M^{\overline p \overline p'}$ with $p=1$ and $p'=1,\dots,4$ and with $p=1,\dots,4$ and $p'=1$, respectively, through (5.8.1) and (5.8.2), $M^{p \underline p'}$ and $M^{ p \overline p'}$ with $p=1,\dots,3$ and $p'=1$ and  $M^{\overline p  p'}$ and $M^{\underline p  p'}$ with $p=1$ and $p'=1,\dots,3$ through (5.8.4). The scalars that remain massless agree with those of a $3$-dimensional toroidal compactification with a $(U(1)\times SU(3))_L\times (U(1)\times SU(3))_R$ enhancement. When truncating away the massive modes, the action (4.7) reduces to the one with $(U(1)\times SU(3))_L\times (U(1)\times SU(3))_R$. After realizing that the structure constants are in block form, {\em i.e.} they do not mix indices from $SU(2)$ and $SU(3)$, it is straightforward to see from (4.8) that the kinetic terms reduce consistently. For instance, $D_{\mu}M_{11}=\partial_{\mu}M_{11}$, $F^1_{\mu\nu}=\partial_{[\mu}A^1_{\nu]}$, etc., while the corresponding expressions for the fields with entries in $SU(3)$ remain as in (4.8) but only with the $SU(3)$ block of structure constants. Concerning the potential term, contributions coming from structure constants with a leg in $SU(2)$ are projected out and a potential compatible with a $(U(1)\times SU(3))_L\times (U(1)\times SU(3))_R$ enhancement is retained.

If instead of $v^{11}$ the only non-vanishing vev is $v^{33}$ (alternatively, $v^{22}$), $SU(2)\times SU(3)$ beaks to $U(1)\times SU(2)^2$. Since $v^{1p'}=v^{2p'}=0$ for any $p'$, it follows that $A^{\underline{1}}$ and $A^{\underline{2}}$ remain massless, and the same happens when exchanging left and right. Vectors with a leg in a specific simple root of $SU(3)$, namely, $A^{\underline{3}}$, $A^{\overline{3}}$, $\overline A^{\underline{3}}$ and $\overline A^{\overline{3}}$, acquire mass through (5.2). Also  $A^{\underline{4}}$, $A^{\overline{4}}$, $\overline A^{\underline{4}}$ and $\overline A^{\overline{4}}$ acquire mass, this time since neither $n_3{}^4$ nor $v^{33}$ vanish. Scalars with at least an entry equal to these specific simple roots or the non-simple roots acquire mass as well due to (5.8). The remaining massless field content is, thus, consistent with a $(U(1)\times SU(2)^2)_L\times(U(1)\times SU(2)^2)_R$ enhancement. When projected out onto the massless sector, the action (4.7) reduces to the one with this symmetry. This time the proof is trickier than before. It follows from (4.8) after realizing that the structure constants with an entry equal to $3$ are non trivial when the other entries are associated with the corresponding simple roots of $SU(3)$ or the non-simple roots, which are precisely those modes that are removed from the massless sector.

We have been able to embed both maximal enhancements of $T^2$, namely, $SU(3)_L\times SU(3)_R$ and $SU(2)_L^2 \times SU(2)_R^2$, into a single maximal enhancement group associated with a compactification on $T^3$. The increase in the torus dimension does not respond to the necessity of encoding any extra massless string winding mode and it is related with the appearance of the also unwanted $U(1)$ factors. This fact suggests a way of consistently removing the extra ``unphysical'' direction, namely, by decompactifying it. While doing so in (4.7), all vectors and scalars with at least an index in $U(1)$ accommodate as new components in the metric and $B$-field of a higher dimensional extended space. This picture turns out to be consistent if this extended space agrees with the original one, namely, if $d'=d-1$. Heuristically, when we borrow a dimension from the extended $d$-dimensional space, we compactify it to get a $3$-dimensional torus and we consider the action associated with the $(SU(2)\times SU(3))_L\times(SU(2)\times SU(3))_R$ enhancement, we get a unified way of describing both maximal enhancements of $T^2$ if, in the end, we decompactify the extra dimension, giving it back to the extended space.

Concerning the choice of $SU(2)\times SU(3)$, it is clear that it was not the only one available. For instance, another rank $3$ choice would be $SU(4)$. And of course there are infinitely many other groups with rank greater than $3$, which in turn would need a higher number of decompactified dimensions. The choice of $SU(2)\times SU(3)$ is the minimal one since among those groups with lowest rank that have $SU(3)$ and $SU(2)^2$ as regular subgroups, it is the one with the lowest dimension. The dimension of this group is the one we expect: by inspecting Tables 2.1 and 2.2 that show the massless vectors at the two points of maximal enhancement, we see that the first massless vector in Table 2.1 is the same as the first vector in Table 2.2, while the second massless vector in Table 2.1 does not appear in the list of massless vectors in Table 2.2. Thus, the number of left-moving vectors that are massless at some point in moduli space is $2 \times 1$ ($A^{\underline 1}$, $A^{\overline 1}$) plus $2 \times 3$ (the second vector in Table 2.1 together with the second and third vectors in Table 2.2) plus the two Cartan vectors which gives a total of 10. This is the same as the dimension of   $SU(2)\times SU(3)$ minus one for the extra Cartan direction that is decompactified at the end of the process.



The discussion for the $2$-dimensional torus can be generalized for higher dimensions. Indeed, the problem of describing in a unified way the entire moduli space associated with a compactification on a $k$-dimensional torus could be solved by looking at a compactification in an enhancement point on a torus of higher dimension, say $k+p$, such that the entire moduli space of the $k$-dimensional torus in the boundary can be reached, after symmetry breaking, by decompactifying $p$ directions. 
These decompactified directions will be absorbed by the extended space so they will remain physical if we ``borrow'' them from the extended space at first: they ``return'' to space-time after the process ends.

This strategy can always be implemented by simply considering the direct product of all maximal enhancement groups, but of course this choice does not constitute the most ``economical'' one. In order to discuss how to get the minimal group, let us briefly recall Dynkin's recipe for obtaining maximal regular subalgebras of a given algebra \cite{DiFrancesco}. Let $\mathfrak g$ be a semi-simple Lie algebra. Its maximal regular subalgebras are constructed either by removing from the associated extended Dynkin diagram a node whose mark is a prime number, this yields the semi-simple ones, or by removing two nodes with mark $1$ and the addition of a $\mathfrak{u}(1)$ factor, which gives the non-semi-simple subalgebras. As we have already seen, maximal regular semi-simple subalgebras cannot be reached by means of a spontaneous symmetry breaking. Therefore, we shall concentrate on non-semi-simple maximal subalgebras.

For simply-laced algebras Dynkin's procedure for obtaining non-semi-simple regular maximal subalgebras simplifies. Indeed, the maximal non-semi-simple regular subalgebras of $\mathfrak{su}(N)$ are obtained by removing a single node from its diagram, while for $\mathfrak{so}(2N)$ they are obtained after removing a boundary node. For the exceptional Lie algebras $\mathfrak{E}_6$ and $\mathfrak{E}_7$, the non-semi-simple regular subalgebras can be determined by removing a boundary node lying in the longest leg of its Dynkin diagram. $\mathfrak{E}_8$, on the other hand, has no non-semi-simple maximal regular subalgebras at all. A spontaneous symmetry breaking is related to a chain of projections that reduce to one of these when restricted to the semi-simple part of the gauge group.

Having all these facts in mind, it is straightforward to visualize the procedure for constructing the minimal enhancement group. Let us work out some specific examples.

For $T^2$ the maximal enhancement points correspond to the groups\footnote{From now on we discuss only one sector of the enhancement group $G \times G$.} $SU(2)^2$ and $SU(3)$, whose Dynkin diagrams are given in Table 6.1.
 The minimal group including these ones regularly must have rank less than or equal to $4$ since the enhancement point for the torus $T^4$ with diagrams $\begin{dynkin} \dynkinline{3}{0}{4}{0}; \foreach \x in {1,...,4} {\dynkindot{\x}{0}} \end{dynkin}$ includes them. By removing a node from this diagram we get two enhancement points for $T^3$ that actually work: $SU(3)\times SU(2)$, with diagram $\begin{dynkin} \dynkinline{2}{0}{3}{0}; \foreach \x in {1,...,3} {\dynkindot{\x}{0}} \end{dynkin}$, and $SU(4)$, whith diagram $\begin{dynkin} \dynkinline{1}{0}{2}{0}; \dynkinline{2}{0}{3}{0}; \foreach \x in {1,...,3} {\dynkindot{\x}{0}} \end{dynkin}$. We recognize the answer in the first group, since it has  the smallest dimension.

\begin{table}[h]
\centering
\label{t2table}
\begin{tabular}{|c|c|c||c|c|c|}
\hline
\multicolumn{3}{|c||}{Enhancement group}  & \multicolumn{3}{c|}{Unifying group} \\
\hline
Rank & Group & Dynkin diagram & Rank & Group & Dynkin diagram  \\
\hline
\multirow{2}{*}{2} & $SU(2)^2$ & $\begin{dynkin} \foreach \x in {1,2} {\dynkindot{\x}{0}} \end{dynkin}$ & \multirow{2}{*}{3} & \multirow{2}{*}{$SU(2)\times SU(3)$} & \multirow{2}{*}{$\begin{dynkin} \dynkinline{2}{0}{3}{0}; \foreach \x in {1,...,3} {\dynkindot{\x}{0}} \end{dynkin}$} \\
                  & $SU(3)$ & $\begin{dynkin} \dynkinline{1}{0}{2}{0}; \foreach \x in {1,2} {\dynkindot{\x}{0}} \end{dynkin}$ &                   &                   &   \\
\hline
\end{tabular}
\caption{The case $T^2$. The groups refer to one single sector, say the left-moving one.}
\end{table}

For $T^3$ the maximal enhancement points are given in Table 6.2.
Again, it is possible to include all these graphs as subdiagrams of the enhancement group of a torus with a single extra dimension. Indeed, $SU(4)\times SU(2)$ and $SU(5)$, whose diagrams are $\begin{dynkin} \dynkinline{1}{0}{2}{0}; \dynkinline{2}{0}{3}{0}; \foreach \x in {1,...,4} {\dynkindot{\x}{0}} \end{dynkin}$ and $\begin{dynkin} \dynkinline{1}{0}{2}{0}; \dynkinline{2}{0}{3}{0}; \dynkinline{3}{0}{4}{0}; \foreach \x in {1,...,4} {\dynkindot{\x}{0}} \end{dynkin}$, do the job. The first one has the lowest dimension.

\begin{table}[h]
\centering
\begin{tabular}{|c|c|c||c|c|c|}
\hline
\multicolumn{3}{|c||}{Enhancement group}  & \multicolumn{3}{c|}{Unifying group} \\
\hline
Rank & Group & Dynkin diagram & Rank & Group & Dynkin diagram  \\
\hline
\multirow{3}{*}{3} & $SU(2)^3$ & $\begin{dynkin} \foreach \x in {1,...,3} {\dynkindot{\x}{0}} \end{dynkin}$ & \multirow{3}{*}{4} & \multirow{3}{*}{$SU(4)\times SU(2)$ } & \multirow{3}{*}{ $\begin{dynkin} \dynkinline{1}{0}{2}{0}; \dynkinline{2}{0}{3}{0}; \foreach \x in {1,...,4} {\dynkindot{\x}{0}} \end{dynkin}$ } \\
                  & $SU(3)\times SU(2)$ & $\begin{dynkin} \dynkinline{1}{0}{2}{0}; \foreach \x in {1,...,3} {\dynkindot{\x}{0}} \end{dynkin}$ &                   &                   &   \\
									& $SU(4)$ & $\begin{dynkin} \dynkinline{1}{0}{2}{0}; \dynkinline{2}{0}{3}{0}; \foreach \x in {1,...,3} {\dynkindot{\x}{0}} \end{dynkin}$ &                   &                   &   \\
\hline
\end{tabular}
\caption{The case $T^3$}
\label{t3table}
\end{table}

The $4$-dimensional torus involves new aspects. The diagrams to ensemble are those listed in Table 6.3.

\begin{table}[h]
\centering
\begin{tabular}{|c|c|c||c|c|c|}
\hline
\multicolumn{3}{|c||}{Enhancement group}  & \multicolumn{3}{c|}{Unifying group} \\
\hline
Rank & Group & Dynkin diagram & Rank & Group & Dynkin diagram  \\
\hline
\multirow{6}{*}{4} & $SU(2)^4$ & $\begin{dynkin} \foreach \x in {1,...,4} {\dynkindot{\x}{0}} \end{dynkin}$ & \multirow{6}{*}{7} & \multirow{6}{*}{$SO(12)\times SU(2)$ } & \multirow{6}{*}{ $\begin{dynkin} \dynkinline{1}{0}{2}{0}; \dynkinline{2}{0}{3}{0}; \dynkinline{2}{0}{2}{1}; \dynkinline{3}{0}{4}{0}; \dynkinline{4}{0}{5}{0}; \dynkindot{2}{1} \foreach \x in {1,...,6} {\dynkindot{\x}{0}} \end{dynkin}$ } \\
                  & $SU(3)\times SU(2)^2$ & $\begin{dynkin} \dynkinline{1}{0}{2}{0}; \foreach \x in {1,...,4} {\dynkindot{\x}{0}} \end{dynkin}$ &                   &                   &   \\
									& $SU(4)\times SU(2)$ & $\begin{dynkin} \dynkinline{1}{0}{2}{0}; \dynkinline{2}{0}{3}{0}; \foreach \x in {1,...,4} {\dynkindot{\x}{0}} \end{dynkin}$ &                   &           &   \\
									& $SU(3)^2$ & $\begin{dynkin} \dynkinline{1}{0}{2}{0}; \dynkinline{3}{0}{4}{0}; \foreach \x in {1,...,4} {\dynkindot{\x}{0}} \end{dynkin}$ & &           &   \\
									& $SU(5)$ & $\begin{dynkin} \dynkinline{1}{0}{2}{0}; \dynkinline{2}{0}{3}{0}; \dynkinline{3}{0}{4}{0}; \foreach \x in {1,...,4} {\dynkindot{\x}{0}} \end{dynkin}$ & &           &   \\
									& $SO(8)$ & $\begin{dynkin} \dynkinline{1}{0}{2}{0}; \dynkinline{2}{0}{3}{0}; \dynkinline{2}{0}{2}{1}; \dynkindot{2}{1} \foreach \x in {1,...,3} {\dynkindot{\x}{0}} \end{dynkin}$ & &           &   \\
\hline
\end{tabular}
\caption{The case $T^4$}
\label{t4table}
\end{table}

A new feature is the fact that one of the maximal enhancement points is associated with a group, $SU(2)^4$, which is a regular subgroup of the group associated with another maximal enhancement point, {\em i.e.} $SO(8)$. Nevertheless, as we have already mentioned, two enhancement points such that the group for one of them is a semi-simple maximal regular subgroup of the other are not linked by a spontaneous symmetry breaking. Both diagrams must be considered separately.

It can straightforwardly be  checked that it is impossible to include all the diagrams in a diagram associated with a torus with a single extra dimension. Indeed, the minimal diagram including the full list in Table \ref{t4table} turns out to be $SO(12)\times SU(2)$, namely, a maximal enhancement point of $T^7$ with diagram $\begin{dynkin} \dynkinline{1}{0}{2}{0}; \dynkinline{2}{0}{3}{0}; \dynkinline{2}{0}{2}{1}; \dynkinline{3}{0}{4}{0}; \dynkinline{4}{0}{5}{0}; \dynkindot{2}{1} \foreach \x in {1,...,6} {\dynkindot{\x}{0}} \end{dynkin}$. (Note that if regular maximal semi-simple subgroups were reached by a symmetry breaking, the minimal diagram would be $SO(12)$, namely, a point in $T^6$)

In the following table we present the unifying scheme for $T^5$. For $k\ge 6$ the new feature is the appearance of the exceptional groups in the list. Since these groups must be treated separately, they do not add more than abstruse combinatorics to the problem and we prefer not to make explicit the solutions beyond $T^5$.

\begin{table}[h]
\centering
\begin{tabular}{|c|c|c||c|c|c|}
\hline
\multicolumn{3}{|c||}{Enhancement group}  & \multicolumn{3}{c|}{Unifying group} \\
\hline
Rank & Group & Dynkin diagram & Rank & Group & Dynkin diagram  \\
\hline
\multirow{9}{*}{5} & $SU(2)^5$ & $\begin{dynkin} \foreach \x in {1,...,5} {\dynkindot{\x}{0}} \end{dynkin}$ & \multirow{9}{*}{9} & \multirow{9}{*}{$SO(16)\times SU(2)$ } & \multirow{9}{*}{ $\begin{dynkin} \dynkinline{1}{0}{2}{0}; \dynkinline{2}{0}{3}{0}; \dynkinline{2}{0}{2}{1}; \dynkinline{3}{0}{4}{0}; \dynkinline{4}{0}{5}{0}; \dynkinline{5}{0}{6}{0}; \dynkinline{6}{0}{7}{0}; \dynkindot{2}{1} \foreach \x in {1,...,8} {\dynkindot{\x}{0}} \end{dynkin}$ } \\
                  & $SU(3)\times SU(2)^3$ & $\begin{dynkin} \dynkinline{1}{0}{2}{0}; \foreach \x in {1,...,5} {\dynkindot{\x}{0}} \end{dynkin}$ &                   &                   &   \\
									& $SU(4)\times SU(2)^2$ & $\begin{dynkin} \dynkinline{1}{0}{2}{0}; \dynkinline{2}{0}{3}{0}; \foreach \x in {1,...,5} {\dynkindot{\x}{0}} \end{dynkin}$ &                   &           &   \\
									& $SU(3)^2\times SU(2)$ & $\begin{dynkin} \dynkinline{1}{0}{2}{0}; \dynkinline{3}{0}{4}{0}; \foreach \x in {1,...,5} {\dynkindot{\x}{0}} \end{dynkin}$ & &           &   \\
									& $SU(5)\times SU(2)$ & $\begin{dynkin} \dynkinline{1}{0}{2}{0}; \dynkinline{2}{0}{3}{0}; \dynkinline{3}{0}{4}{0}; \foreach \x in {1,...,5} {\dynkindot{\x}{0}} \end{dynkin}$ & &           &   \\
									& $SU(4)\times SU(3)$ & $\begin{dynkin} \dynkinline{1}{0}{2}{0}; \dynkinline{2}{0}{3}{0}; \dynkinline{4}{0}{5}{0}; \foreach \x in {1,...,5} {\dynkindot{\x}{0}} \end{dynkin}$ &                   &           &   \\
									& $SU(6)$ & $\begin{dynkin} \dynkinline{1}{0}{2}{0}; \dynkinline{2}{0}{3}{0}; \dynkinline{3}{0}{4}{0}; \dynkinline{4}{0}{5}{0}; \foreach \x in {1,...,5} {\dynkindot{\x}{0}} \end{dynkin}$ & &           &   \\
									& $SO(8) \times SU(2)$ & $\begin{dynkin} \dynkinline{1}{0}{2}{0}; \dynkinline{2}{0}{3}{0}; \dynkinline{2}{0}{2}{1}; \dynkindot{2}{1} \foreach \x in {1,...,4} {\dynkindot{\x}{0}} \end{dynkin}$ & &           &   \\
									& $SO(10)$ & $\begin{dynkin} \dynkinline{1}{0}{2}{0}; \dynkinline{2}{0}{3}{0}; \dynkinline{3}{0}{4}{0}; \dynkinline{2}{0}{2}{1}; \dynkindot{2}{1} \foreach \x in {1,...,4} {\dynkindot{\x}{0}} \end{dynkin}$ & &           &   \\
\hline
\end{tabular}
\caption{The case $T^5$}
\label{t5table}
\end{table}

Let us now check that this formal group theory exercise gives answers with (very) sensible physical meaning. In the $T^2$ case, we have shown that the dimension of the smallest ``unifying group", $SU(2) \times SU(3)$, is the one we expect by counting the number of different left-moving vectors that are massless at either the $SU(2)^2$ or the $SU(3)$ enhancement point. The same exercise for $T^3$ gives a surprising answer: the number of different massless vectors combining those at the three maximal enhancement points is 16. We would thus expect the group that contains all these to have dimension equal or greater than 20 (16 + 3 Cartan directions on  $T^3$  + 1 extra Cartan from the direction to be decompactified). However, the smallest unifying group, shown in Table \ref{t3table}, is $SU(4) \times SU(2)$, which has dimension 18. Is group theory giving a more economical choice that one needs from string theory? The answer is negative, and we can understand the result from both sides. From string theory, we show in Appendix C that by choosing different directions on  $T^4$ to take the decompactification limit, one can recover from the 14 vectors associated to roots of $SU(4) \times SU(2)$, the total of 16 different vectors that are massless at some point in the moduli space of the $T^3$ compactification. From group theory, this is equivalent to saying that a given root on a 4-dimensional lattice can give rise to different 3-dimensional roots by taking different 3-dimensional subspaces. 



\section{Conclusions}
\label{Conc}

 The Abelian gauge symmetries of KK compactifications in point-particle theories correspond to isometries of the compact directions. In string theory,  the appearance of  winding modes  in addition to KK modes, suggests that the corresponding isometries would be those of a double torus incorporating the perspective of both, momentum and winding modes. Actually, $U(1)^k_L\times U(1)^k_R$ is the isometry of $T^{2k}$,   and double field theory seems to be the appropriate framework to deal with these double geometries. The $O(d+k,d+k)$ global symmetry of DFT, where $d$ is the number of non-compact dimensions,
allows to symmetrically include a torus and its T-dual one.

 However, the non-Abelian gauge symmetries arising in toroidal compactifications of string theory require additional structures. As shown in \cite{uscircle},  the enhancement of the gauge symmetry occuring in compactifications of one dimension on a string-size circle, requires to promote the $O(d+1,d+1)$ symmetry to $O(d+3, d+3)$.
Now, we have incorporated the enhanced gauge symmetry  arising at special points in the  moduli space of compactifications on $T^k$
 by building an $O(d+n, d+n)$ structure, where  $n+n$ is the dimension of the left and right enhanced gauge groups.  The effective action of toroidal compactifications of  bosonic string theory was reproduced by a generalized Scherk-Schwarz reduction of the DFT action. The reduced action gives the precise string theory masses of vectors and scalars close to a point of maximal enhancement, which is nicely described in the effective action by spontaneous symmetry breaking.  We have also shown that the full sector of states that become massless at any point in moduli space can be described by considering a compactification on a higher-dimensional torus at a certain point of maximal enhancement.

We have presented two possible expressions for the twist matrix realizing the enhanced gauge algebra.  
In the realization constructed in section \ref{sec:internal}, 
one can interpret the vector fields with non-vanishing compact momentum or winding number as arising from a  Kaluza-Klein compactification of a ``metric" and ``$B$"-field living on a $2n$-dimensional manifold, $n+n$ being the dimension of the $G\times G$ enhanced gauge group, and the twist matrix as the corresponding generalized vielbein.  This generalized vielbein depends on the coordinates of a double $k$-dimensional torus associated to the Cartan generators. The structure constants of the gauge algebra can be obtained from this vielbein using the deformed C-bracket (\ref{deform}), which is consistent with the local symmetries and the $O(k,k)$ covariance of the theory. The deformation accounts for the cocycle factors that are necessary in the current algebra representation and breaks the $O(d+n, d+n)$ symmetry to $O(d+k, d+k)$. The ``cocycle tensor" used to deform the algebra reproduces the structure constants involving only ladder generators, to which we have not associated any coordinate, and thus it satisfies the last consistency constraint in \eqref{constraints}. This deformation is reminiscent of the deformation of the bracket needed to account for Romans mass in massive type IIA supergravity \cite{Romans}.  

An interesting property of this generalized vielbein is that it is an eigenvector of  $\partial_M\partial^M$. Actually, it satisfies a modified version of the weak constraint (holding even when $N\ne \bar N$) which looks like the operator form of the level matching condition accounting for the extra massless fields arising at the enhancement point. This observation clarifies a recurrent discussion about the consistency of generalized Scherk-Schwarz compactifications, which were thought to violate the level matching condition. Now we see that the weak constraint is modified  and it agrees with the level matching condition.

The alternative construction of the vielbein in section \ref{sec:vielbeingroupmanifold} is based on the  formulation of DFT on group manifolds \cite{DFTgroup,dh}. In this framework, the vielbein depends on the $n$ coordinates of the group manifold corresponding to the maximally enhanced symmetry group. Although this gives a {\it geometric} frame, in the sense that it does not depend on a double set of  coordinates and then it obeys the strong constraint in $2n$ dimensions, from the point of view of compactifications on a $k$-torus this frame would be non-geometric, as it depends on more than $k$ coordinates. However, the extra $n-k$ coordinates have no clear interpretation  from the torus point of view.

 DFT allows to anticipate the T-duality symmetry of string theory before dimensional reduction, and thus it provides a fruitful framework to better understand this symmetry. Much progress has been achieved since the original ideas were put forward \cite{Siegel:1993xq} and several genuine stringy features have been reproduced within this field theory context. Indeed, it has been possible to go beyond supergravity incorporating $\alpha'$-corrections \cite{dn},  massive string states \cite{amn} and now, the gauge symmetry enhancement of generic toroidal compactifications. Nevertheless, the geometry of the double space remains elusive and more work is necessary to comprehend and grasp the significance of the $O(D,D)$ geometry underlying DFT.

\vskip 1cm


\subsection*{Acknowledgements}
We thank C. Hull, G. Inverso, D. Marques, C. Strickland Constable, D. Waldram and B. Zwiebach for valuable insights.
This work was partially supported
by EPLANET, CONICET-PIP 2015-2017, UBACyT 2014-2017 and the ERC Starting Independent Researcher
Grant 259133-ObservableString.


\appendix


\section{Lie Algebras }
\label{app:Liealgebras}

In this Appendix we collect the necessary definitions and notation used in the text regarding simply-laced Lie algebras of dimension $n$. We follow closely \cite{DiFrancesco}. We start with a general discussion and then specialize to the groups needed in the text, namely $SU(2)^2$, $SU(3)$ and $SU(2) \times SU(3)$. 

A Lie algebra can be specified by a set of generators ${X^a}$, $a=1,...,n$ and their commutation relations
\beq
\left[X^a,X^b\right]=f^{ab}{}_c X^c
\eeq
where the constants $f^{ab}{}_c$ are the structure constants.
In the standard Cartan-Weyl basis, the Lie generators are defined as follows. First, a set of Cartan generators $H^m$ ($m=1,\dots,k$) expanding a maximal commutative subalgebra (the so-called Cartan subalgebra) is fixed. Here $k$, the dimension of the Cartan subalgebra, stands for the rank of the full Lie algebra. The remaining  generators are chosen to be the simultaneous eigenvectors of all $H^m$ under the adjoint action, namely, satisfying
\begin{equation}
[H^m,J^{\alpha}]=\alpha^m J^{\alpha}.
\end{equation}
The $k$-dimensional vector $\alpha$ is called a root of the Lie algebra, and the associated generator $J^{\alpha}$ a "ladder" operator. Since the Cartan subalgebra is maximal, roots are non-degenerate. Notice that Hermitian conjugation of the previous equation shows that $-\alpha$ is a root whenever $\alpha$ is, provided $J^{-\alpha}=J^{\alpha\dag}$.

The Jacobi identity shows that the commutator of two ladder generators associated with opposite roots commute with all the Cartan generators and therefore it must be an element of the Cartan subalgebra as well. The choice of this element determines the normalization of ladder operators. Usually it is fixed as\footnote{Some of the equations are written using the fact that for simply-laced Lie algebras $|\alpha|^2=\alpha^m \delta_{mn}\alpha^n=2$.}
\begin{equation}
[J^{\alpha},J^{-\alpha}]=\alpha\cdot H,
\end{equation}
where $\alpha \cdot H=\alpha^m \delta_{mn} H^n$. On the other hand, the commutator of two  ladder generators $J^{\alpha}$ and $J^{\beta}$ must vanish if $\alpha + \beta$ is not a root and it must be proportional to $J^{\alpha + \beta}$ otherwise, explicitly:
\begin{equation}
[J^{\alpha},J^{\beta}]=\cal N_{\alpha,\beta}J^{\alpha + \beta}\qquad \text{if $\alpha + \beta$ is a root. }
\label{eq:Nab}
\end{equation}

The Killing form is defined as
\begin{equation}
K(X,Y)=\frac{1}{2g}{\mbox{Tr}}\left({\mbox{Ad}X\mbox{Ad}Y}\right),
\end{equation}
where $g$ is the dual Coxeter number of the Lie algebra. In the Cartan-Weyl basis (and for a simply-laced algebra) we have
\begin{equation} \label{KillingH}
K(H^{m},H^{n})=\delta^{mn}, \qquad K(J^{\alpha},J^{-\alpha})=1, 
\end{equation}
for any root $\alpha$.

Any root $\alpha$ can be written as 
\begin{eqnarray}
\alpha=n_m \beta_m
\label{eq:}
\end{eqnarray}
where $\left\{\beta_m\right\}$ is a basis for the $k$-dimensional space of roots. The root $\alpha$ is said to be positive if the first non-zero number in the sequence $(n_1,n_2,...)$ is positive and, consequently, it will be associated to a raising operator. 
We will be using the following notation in the remainder of this work: taking the set $\left\{ \alpha_i \right\}$ of positive roots, $i=1,...,\tfrac12 (n-k)$, we will denote the raising and lowering ladder operators as: 
\begin{eqnarray}
J^{\underline i}&=&J^{\alpha_i} \qquad \ \, \text{for raising operators} \nn\\
J^{\overline \imath}&=&J^{-\alpha_i} \qquad \text{for lowering operators}
\end{eqnarray}
The structure constants will have the same type of index, for example (no summation over $\underline i$):
\beq
\left[H^m,J^{\underline i}\right]= f^{m\underline i}{}_{\underline i} J^{\underline i} \ , \quad {\rm with} \ f^{m\underline i}{}_{\underline i}=\alpha_i{}^m
\eeq
where $\alpha_i{}^m=(\alpha_i)^m$ is the $m$-th component of the vector $\alpha_i$. 

Simple roots are those that cannot be written as a combination of two positive roots. They provide a suitable basis for the dual space of the Cartan subalgebra. We shall denote by $\alpha_p$ with $p=1,\dots,k$ the simple roots. It is a classical result that the full set of roots of a given semisimple Lie algebra can be reconstructed through the action of the Weyl group from the set of simple roots, which in turn can be straightforwardly extracted from the so-called Cartan matrix 
\beq
A^{pq}=\alpha_p\cdot\alpha_q=\alpha_p{}^m \delta_{mn}\alpha_q{}^{n} .
\eeq
Moreover, this matrix completely fixes the commutation relations of the algebra as can be seen from the Chevalley basis, to which we now turn.

In the Chevalley basis, to each simple root $\alpha_p$ there corresponds two generators $e^p$ and $f^p$, associated, respectively, with the raising operator $J^{\underline p}$ and the lowering operator $J^{\overline p}$. The Cartan generators are defined as
\begin{equation} \label{defh}
h^p=\alpha_p\cdot H.
\end{equation}

Their commutation relations read
\begin{eqnarray}
\label{commchevalley}
&& [h^p,h^q]=0, \nonumber \\
&& [h^p,e^q]=A^{qp} \, e^q, \nonumber \\
&& [h^p,f^q]=-A^{qp} \, f^q, \nonumber \\
&& [e^p,f^q]=\delta^{pq} \, h^q, 
\end{eqnarray}
where no summation over $p$ or $q$ is understood on the right hand sides.

 The remaining ladder generators (recall that there are $\tfrac12(n-3k)$ positive non-simple roots) are obtained by repeated commutation of these ones, subject to the so-called Serre relations:
\begin{equation}
\label{serre}
\left(\mbox{Ad}(e^p)\right)^{1-A^{qp}}e^q=\underbrace{\left[e^p,\left[e^p,\cdots,\left[e^p\right.\right.\right.}_{(1-A^{qp}) \, \mbox{\tiny times}}\left.\left.\left.,e^q\right]\right]\right]=0,
\end{equation}
and similarly for $f$. 

The non-simple roots are a linear combination of the simple roots 
\beq
\alpha_u= n_p{}^u \alpha_p 
\eeq
where $u=1,...,\tfrac12(n-3k)$ labels non-simple roots. If $\alpha_p$ are positive roots, and all $n_p$ are positive for a given root $u$, then $\alpha_u$ is a positive root. 
 We shall write the commutation relations between ladder operators associated to non-simple roots and the Cartan subalgebra as (no summation over $u$)
\beq
\left[h^p,e^{u}\right]=c^{pu} e^{u} , \nn
\label{eq:cui}
\eeq
\beq
\left[h^p,f^{u}\right]=-c^{pu} f^{u} ,
\label{eq:cui2}
\eeq
where
\beq \label{c} 
c^{pu}=\alpha_{p} \cdot \alpha_u = A^{pq} {} n_q{}^u .
\eeq 
We thus get
\beq
\left[e^u,f^u \right]=n_p{}^u \, h^p \ .  
\eeq

The Killing form in the Chevalley basis is given by
\begin{equation} \label{KillinghatH}
K(e^i,f^j)=\delta^{ij}, \qquad K(h^p,h^q)=A^{pq}. 
\end{equation}

It will be useful to write  explicit expressions for some $f_{a b c}$ (all indices down) as there are many contractions that involve structure constants with indices in this position.
The $p-th$ row of the Cartan matrix gives the structure constants in the Chevalley basis for the $p-th$ simple root, $i.e.$:
\beq
A^{pq}=f^{q\underline p}{}_{\underline p},
\eeq
and lowering all indices with $K^{-1}$ we get
\beq
f_{q\overline p\underline p}=
\delta_{pq},
\label{eq:simple}
\eeq
which takes a very simple form.
For non-simple roots we have (no summation over $\underline u$)
\beq
f^{p\underline u}{}_{\underline u}=c^{pu},
\eeq
and then
\beq
f_{p\overline u\underline u}=n_{p}{}^u.
\label{eq:nonsimple}
\eeq

\subsection{$SU(2)^2$}

For $SU(2)^2$ the simple roots can be taken as 
\beq
\alpha_1=\sqrt 2(1,0) \ , \qquad \alpha_2=\sqrt 2(0,1) \ .
\eeq

The Cartan matrix is
\begin{equation} \label{ASU22}
A=\left(\begin{array}{cc}
	2 & 0 \\
	0 & 2
\end{array}\right).
\end{equation}
The non-trivial commutation relations in the Chevalley basis read
\begin{eqnarray}
&& [h^1,e^1]=2e^1, \qquad [e^1,f^1]=h^1, \qquad [h^1,f^1]=-2f^1, \nonumber \\
&& [h^2,e^2]=2e^2, \qquad [e^2,f^2]=h^2, \qquad [h^2,f^2]=-2f^2\, ,
\end{eqnarray}
and the (antisymmetric) structure constants are given by
\begin{equation}
f_{1\overline 1\underline 1}=f_{2\overline 2\underline 2}=1,
\end{equation}
all the other being zero.

\subsection{$SU(3)$}
\label{app:SU3}
For $SU(3)$ we take the simple positive roots to be  
\beq \label{simplerootssu3}
\alpha_1=\sqrt{2}\left(\frac{\sqrt{3}}{2},-\frac{1}{2}\right)  \ , \qquad \alpha_2=\sqrt{2}\left(0,1\right)\ ,
\eeq
and the Cartan matrix is
\begin{equation} \label{ASU3}
A=\left(\begin{array}{cc}
	2 & -1 \\
	-1 & 2
\end{array}\right) \ . 
\end{equation}
Therefore, the commutation relations in the Chevalley basis read
\begin{eqnarray}
&& [h^1,e^1]=2e^1, \qquad [e^1,f^1]=h^1, \qquad [h^1,f^1]=-2f^1, \nonumber \\
&& [h^2,e^2]=2e^2, \qquad [e^2,f^2]=h^2, \qquad [h^2,f^2]=-2f^2, \nonumber \\
&& [h^1,e^2]=-e^2, \qquad [h^1,f^2]=f^2, \nonumber \\
&& [h^2,e^1]=-e^1, \qquad [h^2,f^1]=f^1¬, .
\end{eqnarray}
After introducing
\begin{eqnarray}
&& e^3=[e^1,e^2],  \\
&& f^3=-[f^1,f^2], \nonumber
\end{eqnarray}
the Serre relations reduce to
\begin{eqnarray}
&& [e^1,e^3]=[e^2,e^3]=0, \\
&& [f^1,f^3]=[f^2,f^3]=0. \nonumber
\end{eqnarray}
Once these relations are used, we get the remaining non-trivial commutation formulas, namely
\begin{eqnarray}
&& [h^1,e^3]=e^3, \qquad [h^1,f^3]=-f^3, \nonumber \\
&& [h^2,e^3]=e^3, \qquad [h^2,f^3]=-f^3, \nonumber \\
&& [e^3,f^1]=-e^2, \qquad [e^1,f^3]=-f^2, \label{noncartan}\\
&& [e^3,f^2]=e^1, \qquad [e^2,f^3]=f^1, \nonumber \\
&& [e^3,f^3]=h^1+h^2 \, ,\nn
\end{eqnarray}
and thus the $2 \times 1$ matrix $c^{pu}$ defined in \eqref{c} is $c^{13}=c^{23}=1$.

The (antisymmetric) structure constants are
\begin{eqnarray}
&& f_{1\overline 1\underline 1}=f_{1\overline 3\underline 3}=f_{2\overline 2\underline 2}=f_{2\overline 3 \underline 3}=1, \nonumber \\
&& f_{\overline 1\overline 2\underline 3} =-f_{\underline 1\underline 2\overline 3}=1. 
\end{eqnarray}
all the other being zero.

\subsection{$SU(2) \times SU(3)$}

For $SU(2) \times SU(3)$ we use 
\beq
\alpha_1=\sqrt{2}(1,0,0) \ , \quad  \alpha_2=\sqrt{2}\left(0,\frac{\sqrt{3}}{2},-\frac{1}{2}\right)  \ , \qquad \alpha_3=\sqrt{2}\left(0,0,1\right)\ . 
\eeq
The Cartan matrix is
\begin{equation}
A=\left(\begin{array}{ccc}
  2 & 0 & 0 \\
	0 & 2 & -1 \\
	0 & -1 & 2
\end{array}\right)\, ,
\label{cartansu2su3}
\end{equation}
and therefore,
\begin{eqnarray}
&& [h_1,e_1]=2e_1, \qquad [e_1,f_1]=h_1, \qquad [h_1,f_1]=-2f_1, \nonumber \\
&& [h_2,e_2]=2e_2, \qquad [e_2,f_2]=h_2, \qquad [h_2,f_2]=-2f_2, \nonumber \\
&& [h_3,e_3]=2e_3, \qquad [e_3,f_3]=h_3, \qquad [h_3,f_3]=-2f_3, \nonumber \\
&& [h_2,e_3]=-e_3, \qquad [h_2,f_3]=f_3, \nonumber \\
&& [h_3,e_2]=-e_2, \qquad [h_3,f_2]=f_2\, .
\end{eqnarray}
After introducing
\begin{eqnarray}
&& e_4=[e_2,e_3], \nonumber \\
&& f_4=-[f_2,f_3], \nonumber
\end{eqnarray}
the Serre relations read
\begin{eqnarray}
&& [e_1,e_2]=[e_1,e_3]=0, \nonumber \\
&& [e_2,e_4]=[e_3,e_4]=0, \nonumber \\
&& [f_1,f_2]=[f_1,f_3]=0, \nonumber \\
&& [f_2,f_4]=[f_3,f_4]=0. 
\end{eqnarray}
Once these relations are used, we get the remaining non-trivial commutation formulas, namely
\begin{eqnarray}
&& [h_2,e_4]=e_4, \qquad [h_2,f_4]=-f_4, \nonumber \\
&& [h_3,e_4]=e_4, \qquad [h_3,f_4]=-f_4, \nonumber \\
&& [e_4,f_2]=-e_3, \qquad [e_2,f_4]=-f_3, \nn\\
&& [e_4,f_3]=e_2, \qquad [e_3,f_4]=f_2, \nonumber \\
&& [e_4,f_4]=h_2+h_3. 
\end{eqnarray}

The Killing form is defined as
\begin{equation}
K(e_i,f_j)=\delta_{ij}, \qquad K(h_i,h_j)=A_{ij}. 
\end{equation}

The structure constants are
\begin{eqnarray}
&& f_{1\overline 1\underline 1}=1, \nonumber \\
&& f_{2\overline 2\underline 2}=f_{2\overline 4\underline 4}=f_{3\overline 3\underline 3}=f_{3\overline 4\underline 4}=1, \nonumber \\
&& f_{\overline 2\overline 3\underline 4}=-f_{\underline 2\underline 3\overline 4}=1. 
\end{eqnarray}
all the other being zero.

\section{Cocycles}
\label{app:cocycles}

 Eq. \eqref{vertexroots} involves the cocyle factors $c_{\alpha}$. These need to be introduced as the OPE of two currents $\tilde J^{\alpha}, \tilde J^{\beta}$ 
\bea \label{app1}
\tilde J^\alpha (z) \tilde J^\beta(w)\sim \frac{\tilde J^{\alpha+\beta}}{z-w} + ...
\eea
should be invariant under the interchange $\tilde J^\alpha(z)\leftrightarrow \tilde J^\beta(w)$ and $z\leftrightarrow w$, while \eqref{app1} picks up a minus sign. The cocycle factors  compensate for the extra sign, i.e.
\bea
J^\alpha (z)= c_\alpha \tilde J^\alpha (z)\, .
\eea
Explicitly, if the mode expansion of $y^i(z)$ is
\bea 
y^i(z)=y_0^i-ip^i\ln z+i\sum_{n\ne 0}\frac{a_n^i}n z^{-n}\, ,
\eea
where in particular
\bea
\left [p^i,y_0^j\right ]=-i\delta_{ij}\, ,
\eea
then $c_\alpha=c_\alpha(p)$. As a result, when $c_\alpha(p)$ passes over a vertex operator, its argument is shifted:
\bea
e^{i\alpha\cdot y}c_\beta(p)=(-1)^{(\alpha ,\beta)}c_\beta(p-\alpha)e^{i\alpha\cdot y}\, .
\eea
To have all signs right in the OPE, we thus need
\bea
c_\alpha (p)c_\beta(p-\alpha)=(-1)^{(\alpha, \beta)}c_\beta(p)c_\alpha(p-\beta)\, .
\eea
Furthermore, a closed algebra requires
\bea
c_\alpha (p)c_\beta(p-\alpha)=\epsilon(\alpha, \beta) c_{\alpha +\beta}(p)\, ,
\eea
with $\epsilon (\alpha, \beta)=\pm 1$. A solution to this equation is given by the following 
\beq 
c_{\alpha} (p)=(-1)^{p \ast \alpha} \ , 
\eeq
where the * product between two elements of the root lattice
\beq
 p=\sum n_i \alpha_i \ , \quad \alpha=\sum m_i \alpha_i \ , \quad n_i \ , m_i \in {\mathbb Z}\, ,
\eeq
is given by
\beq \label{ast}
p  \ast \alpha=\sum_{i>j} n_i m_j (\alpha_i,\alpha_j) \ .
\eeq

\section{Symmetry breaking for the $T^4$}
\label{app:t4t3}
 In this Appendix we explicitly work out the unified description on $T^4$ of the full  moduli space of $T^3$.  Indeed, there are three points of maximal enhancement  in $T^3$, which give rise to the (left) gauge groups $SU(4)$, $SU(3)\times SU(2)$ and $\left(SU(2)\right)^3$. As discussed in the main text, these can be described in terms of the $SU(4)\times SU(2)$ enhancement point of  $T^4$.

Aside from the Cartan vectors, there are in total 16 distinct (in terms of the quantum numbers $w^i$ and $n_i$) raising/lowering vectors if the three enhancement points of $T^3$ are considered together, while in  $SU(4)\times SU(2)$ there are only 14. We will show explicitly that these 14 vectors suffice to account for the 16 distinct ones on $T^3$.

The $SU(4)\times SU(2)$ raising/lowering vectors are given in table \ref{ta:su4su2}.
\begin{table}[h!]
\begin{center}
\begin{tabular}{|c|cccc|cccc|}
\hline	
root  & $w^1$ & $w^2$ & $w^3$ & $w^4$ & $n_1$ & $n_2$ & $n_3$ & $n_4$ \\ \hline
$\alpha_1$ & 1 & 0 & 0 & 0 & 1 & 0 & 0 & 0 \\
$\alpha_2$ & 0 & 1 & 0 & 0 & -1 & 1 & 0 & 0 \\
$\alpha_3$ & 0 & 0 & 1 & 0 & 0 & -1 & 1 & 0 \\
$\alpha_4$ & 0 & 0 & 0 & 1 & 0 & 0 & 0 & 1 \\
$\alpha_5$ & 1 & 1 & 0 & 0 & 0 & 1 & 0 & 0 \\
$\alpha_6$ & 0 & 1 & 1 & 0 & -1 & 0 & 1 & 0 \\
$\alpha_7$ & 1 & 1 & 1 & 0 & 0 & 0 & 1 & 0 \\
$\alpha_{-1}$ & -1 & 0 & 0 & 0 & -1 & 0 & 0 & 0 \\
$\alpha_{-2}$ & 0 & -1 & 0 & 0 & 1 & -1 & 0 & 0 \\
$\alpha_{-3}$ & 0 & 0 & -1 & 0 & 0 & 1 & -1 & 0 \\
$\alpha_{-4}$ & 0 & 0 & 0 & -1 & 0 & 0 & 0 & -1 \\
$\alpha_{-5}$& 0 & -1 & -1 & 0 & 1 & 0 & -1 & 0 \\
$\alpha_{-6}$ & -1 & -1 & 0 & 0 & 0 & -1 & 0 & 0 \\
$\alpha_{-7}$ & -1 & -1 & -1 & 0 & 0 & 0 & -1 & 0 \\
\hline
\end{tabular}
\caption{Momentum and winding numbers  of the massless vectors at the $\left[SU(4) \times SU(2)\right]_L\times  \left[SU(4) \times SU(2)\right]_R$ enhancement point of $T^4$.}
\label{ta:su4su2}
\end{center}
\end{table}

To get the $SU(4)$ gauge group, one gives vevs to $M^{4p},\,M^{p4}$, $p=1,2,3,4$ in which case the vectors associated to $\alpha_{\pm 4}$ acquire mass and should be discarded. The remaining vectors are those from $SU(4)$, as expected (cf. table \ref{ta:su4}). The identification of coordinates between the $T^4$ and the $T^3$ is $x_{T^4}^1\rightarrow x_{T^3}^1,\; x_{T^4}^2\rightarrow x_{T^3}^2,\; x_{T^4}^3\rightarrow x_{T^3}^3$, and $x_{T^4}^4$ corresponds to the decompactified coordinate.
\begin{table}[h!]
\begin{center}
\begin{tabular}{|c|cccc|cccc|}
\hline	
root  & $w^1$ & $w^2$ & $w^3$ & $w^4$ & $n_1$ & $n_2$ & $n_3$ & $n_4$ \\ \hline
$\alpha_1$ & 1 & 0 & 0 & 0 & 1 & 0 & 0 & 0 \\
$\alpha_2$ & 0 & 1 & 0 & 0 & -1 & 1 & 0 & 0 \\
$\alpha_3$ & 0 & 0 & 1 & 0 & 0 & -1 & 1 & 0 \\
$\alpha_4$ & 1 & 1 & 0 & 0 & 0 & 1 & 0 & 0 \\
$\alpha_5$ & 0 & 1 & 1 & 0 & -1 & 0 & 1 & 0 \\
$\alpha_6$ & 1 & 1 & 1 & 0 & 0 & 0 & 1 & 0 \\
$\alpha_{-1}$ & -1 & 0 & 0 & 0 & -1 & 0 & 0 & 0 \\
$\alpha_{-2}$ & 0 & -1 & 0 & 0 & 1 & -1 & 0 & 0 \\
$\alpha_{-3}$ & 0 & 0 & -1 & 0 & 0 & 1 & -1 & 0 \\
$\alpha_{-4}$& 0 & -1 & -1 & 0 & 1 & 0 & -1 & 0 \\
$\alpha_{-5}$ & -1 & -1 & 0 & 0 & 0 & -1 & 0 & 0 \\
$\alpha_{-6}$ & -1 & -1 & -1 & 0 & 0 & 0 & -1 & 0 \\
\hline
\end{tabular}
\caption{Momentum and winding numbers  of the massless vectors at the $SU(4)_L\times SU(4)_R$ enhancement point of $T^3$.}
\label{ta:su4}
\end{center}
\end{table}

The simplest way to understand how to get the $SU(3)\times SU(2)$  gauge group is to look at the simple roots of $SU(4)\times SU(2)$ ($\alpha_p, p={1,2,3,4}$) and try to get those of $SU(3)\times SU(2)$ ($\alpha_p$, $p={1,2,3}$ in table \ref{ta:su3su2}). Looking at the quantum numbers $w^p$ and $n_p$ this is achieved if we discard $\alpha_{3}$ from the $SU(4)\times SU(2)$ and identify the remaining three with the ones of $SU(3)\times SU(2)$. In terms of the vectors, according to \eqref{Mvectors} we must give vevs to $M^{3p}$ and $M^{p3}$ in order to give mass to the vectors associated to $\alpha_{3}$ (this means the vectors associated to the simple root $\alpha_{3}$ and the ones associated to non-simple roots constructed from $\alpha_{3},$ such as $\alpha_i, i={\pm 6,\pm 7}$). The rest of the vectors remain massless and are identified with the ones of $SU(3)\times SU(2)$. The coordinates of the two tori are identified as $x_{T^4}^1\rightarrow x_{T^3}^1,\; x_{T^4}^2\rightarrow x_{T^3}^2,\; x_{T^4}^4\rightarrow x_{T^3}^3$, and $x_{T^4}^3$ corresponds to the decompactified coordinate.

\begin{table}[h!]
\begin{center}
\begin{tabular}{|c|ccc|ccc|}
\hline	
root  & $w^1$ & $w^2$ & $w^3$ & $n_1$ & $n_2$ & $n_3$ 	\\ \hline
$\alpha_1$ & 1 & 0 & 0 & 1 & 0 & 0  \\
$\alpha_2$ & 0 & 1 & 0 & -1 & 1 & 0  \\
$\alpha_3$ & 1 & 1 & 0  & 0 & 1 & 0  \\
$\alpha_4$ & 0 & 0 & 1 & 0 & 0 & 1 \\
$\alpha_{-1}$ & -1 & 0 & 0 & -1 & 0 & 0  \\
$\alpha_{-2}$ & 0 & -1 & 0 & 1 & -1 & 0  \\
$\alpha_{-3}$ & -1 & -1 & 0 & 0 & -1 & 0  \\
$\alpha_{-4}$ & 0  & 0 & -1 & 0 & 0& -1 \\
\hline
\end{tabular}
\caption{Momentum and winding numbers  of the massless vectors at the $[SU(3) \times SU(2)]_L\times [SU(3) \times SU(2)]_R$ enhancement point of $T^3$.}
\label{ta:su3su2}
\end{center}
\end{table}

To get the  $SU(2)^3$ gauge group we must first redefine the lattice of  $T^4$ in order to identify the vectors with the ones in table \ref{ta:su2su2su2} .
\begin{table}[h!]
\begin{center}
\begin{tabular}{|c|ccc|ccc|}
\hline	
 root &  $w^1$ & $w^2$ & $w^3$ & $n_1$ & $n_2$ & $n_3$ \\ \hline
 $\alpha_1$ & 1 & 0 & 0 & 1 & 0 & 0 \\
$\alpha_2$& 0 & 1 & 0 & 0 & 1 & 0 \\
$\alpha_3$ & 0 & 0 & 1 & 0 & 0 & 1 \\
 $\alpha_{-1}$ & -1 & 0 & 0 & -1 & 0 & 0 \\
 $\alpha_{-2}$ & 0 & -1 & 0 & 0 & -1 & 0 \\
 $\alpha_{-3}$ & 0 & 0 & -1 & 0 & 0 & -1 \\
\hline
\end{tabular}
\caption{Momentum and winding numbers  of the massless vectors at the $SU(2)^3_L\times SU(2)^3_R$ enhancement point of $T^3$.}
 \label{ta:su2su2su2}
\end{center}
\end{table}
\begin{table}[h!]
\begin{center}
\begin{tabular}{|c|cccc|cccc|}
\hline	
root & $w^{'1}$ & $w^{'2}$ & $w^{'3}$ & $w^{'4}$ & $n'_1$ & $n'_2$ & $n'_3$ & $n'_4$ \\ \hline
$\alpha_1$ & 1 & 0 & 0 & 0 & 1 & 0 & 0 & 0 \\
$\alpha_2$ & 0 & 1 & -1 & 0 & -1 & 1 & 0 & 0 \\
$\alpha_3$ & 0 & 0 & 1 & 0 & 0 & 0 & 1 & 0 \\
$\alpha_4$  & 0 & 0 & 0 & 1 & 0 & 0 & 0 & 1 \\
$\alpha_5$ & 1 & 1 & -1 & 0 & 0 & 1 & 0 & 0 \\
$\alpha_6$ & 0 & 1 & 0 & 0 & -1 & 1 & 1 & 0 \\
$\alpha_7$ & 1 & 1 & 0 & 0 & 0 & 1 & 1 & 0 \\
$\alpha_{-1}$ & -1 & 0 & 0 & 0 & -1 & 0 & 0 & 0 \\
$\alpha_{-2}$ & 0 & -1 & 1 & 0 & 1 & -1 & 0 & 0 \\
$\alpha_{-3}$ & 0 & 0 & -1 & 0 & 0 & 0 & -1 & 0 \\
$\alpha_{-4}$ & 0 & 0 & 0 & -1 & 0 & 0 & 0 & -1 \\
$\alpha_{-5}$ & 0 & -1 & 0 & 0 & 1 & -1 & -1 & 0 \\
$\alpha_{-6}$ & -1 & -1 & 1 & 0 & 0 & -1 & 0 & 0 \\
$\alpha_{-7}$ & -1 & -1 & 0 & 0 & 0 & -1 & -1 & 0 \\
\hline
\end{tabular}
\caption{Momentum and winding numbers  of the massless vectors at the $[SU(4) \times SU(2)]_L\times [SU(4) \times SU(2)]_R $ enhancement point of $T^4$, in the basis rotated by $T$. }
\label{ta:su4su2prima}
\end{center}
\end{table}
 We define the matrix 
\beq
T=\left(
\begin{array}{cccc}
 1 & 0 & 0 & 0 \\
 0 & 1 & 0 & 0 \\
 0 & -1 & 1 & 0 \\
 0 & 0 & 0 & 1 \\
\end{array}
\right)\, ,
\eeq
such that the winding numbers change as $w'=T w$ and the momentum numbers as $n'=(T^t)^{-1} n$. The new quantum numbers are presented in table \ref{ta:su4su2prima} and we must get rid of the vectors associated to $\alpha'_i, i={\pm 2,\pm 5,\pm 6,\pm 7}$ and keep all the other vectors, the ones associated to $\alpha'_i, i={\pm 1,\pm 3,\pm 4}$ which have the same quantum numbers as the ones coming from $SU(2)\times SU(2)\times SU(2)$ once we identify $x_{T^4}^{'1}\rightarrow x_{T^3}^1,\;x_{T^4}^{'3}\rightarrow x_{T^3}^2, \; x_{T^4}^{'4}\rightarrow x_{T^3}^3$. 

To give masses to the vectors associated to $\alpha'_i, i={\pm 2,\pm 5,\pm 6,\pm 7}$, it suffices to concentrate on $\alpha'_{2}$ as all other roots are linear combinations that include $\alpha'_{2}$, so they will acquire mass if $\alpha'_{2}$ does. This is achieved by giving vevs to $M^{'2p}$ and $M^{'p2}$ according to \eqref{Mvectors}. The coordinate that  will be decompactified is $x_{T^4}^{'2}$ and the rest are identified as $x_{T^4}^{'1}\rightarrow x_{T^3}^1,\; x_{T^4}^{'3}\rightarrow x_{T^3}^2,\; x_{T^4}^4\rightarrow x_{T^3}^3$.

Thus, we see that  the $14$ raising/lowering vectors of  $T^4$ are sufficient to describe all raising/lowering vectors of  $T^3$.


\end{document}